\documentclass[%
 reprint,
 amsmath,amssymb,
 aps,
]{revtex4-2}

\usepackage{graphicx}
\usepackage{dcolumn}
\usepackage{bm}
\usepackage{xcolor} 
\usepackage[colorlinks=true,citecolor=red,filecolor=green,linkcolor=blue]{hyperref}

\begin{document}

\title{A hybrid dynamical-stochastic model of maximum temperature time series of Imphal, Northeast India incorporating nonlinear feedback and noise diagnostics}

\author{
Mairembam Kelvin Singh$^{1}$,
Athokpam Langlen Chanu$^{2,3}$,
R.~K.~Brojen Singh$^{4}$,
Moirangthem Shubhakanta Singh$^{1,*}$%
}

\affiliation{$^{1}$Department of Physics, Manipur University, Canchipur - 795003, Imphal, Manipur, India}
\affiliation{$^{2}$Asia Pacific Center for Theoretical Physics, Pohang, 37673, Republic of Korea}
\affiliation{$^{3}$Department of Physics, Pohang University of Science and Technology (POSTECH), Pohang, 37673, Republic of Korea}
\affiliation{$^{4}$School of Computational \& Integrative Sciences, Jawaharlal Nehru University, New Delhi - 110067, India}

\email{mshubhakanta@yahoo.com}

\def\blue{\textcolor{blue}}
\def\red{\textcolor{red}}

\begin{abstract}
Climate variability is a complex phenomenon resulting from the interaction of numerous components within a climate system across a wide range of temporal and spatial scales. Although significant advances have been made in understanding global climate variability, there are relatively few studies on regional climate modeling, particularly in developing countries. Motivated by this research gap, we perform a systematic analysis of the variability of maximum temperatures recorded for a region called Imphal, which is the capital city of Manipur, located in Northeast India. We develop a data-driven framework for hybrid dynamical-stochastic modeling of maximum temperature data records spanning 73 years. A comprehensive, data-driven quantitative modeling of temperature dynamics for Imphal is not yet available in the literature. Our modeling approach integrates spectral decomposition, nonlinear feedback mechanisms, and stochastic noise characterization within a unified hybrid structure to investigate temperature variability over small and large time scales. Our data-driven approach yields key insights into the temperature dynamics, including a positive temperature increase in the region during the period investigated. Our hybrid model effectively reproduces the observed dynamics of maximum temperature variability of Imphal with high accuracy, validated by robust statistical and nonlinearity tests. The proposed framework offers practical implications for regional climate prediction and risk assessment for Imphal. Additionally, we provide derivations of Langevin and Fokker–Planck equations for the maximum temperature dynamics, offering the theoretical ground and analytical interpretation of the model that links the temperature dynamics with underlying physical principles.\\

\end{abstract}

\maketitle

\section{Introduction}
Climate variability is a complex phenomenon arising from numerous interacting components of a climate system across a wide range of temporal and spatial scales. Among the multiple variables involved in a region's climate dynamics, maximum temperature is particularly important, and studying its variability over long time scales has become a critical area of research, given increasing concerns over global warming~\cite{Perkins20}. Accurate modeling of such variability is essential not only for understanding climate dynamics but also for enhancing the predictability of extreme events and long-term climate projections.

The field of climate modeling has evolved from deterministic formulations to more realistic stochastic and hybrid frameworks. Early models, such as Arrhenius' radiative balance estimates~\cite{Arrhenius96} and the energy balance models of Budyko and Sellers~\cite{Budyko69,Sellers69}, have established the physical basis of climate sensitivity to external forcing within deterministic frameworks. The subsequent development of General Circulation Models (GCMs) by Manabe and Wetherald~\cite{Manabe67} has enabled large-scale numerical simulations of atmospheric and oceanic processes. Lorenz's discovery of deterministic chaos~\cite{Lorenz63}, revealing the intrinsic unpredictability of deterministic origin, has marked a significant turning point in atmospheric systems modeling. Hasselmann's stochastic climate model~\cite{Hasselmann76}, which integrates random atmospheric fluctuations with slow climate processes, has since laid the foundation for modern hybrid dynamical-stochastic frameworks. This has enabled subsequent research on the nonlinear and noise-driven phenomena such as the El Niño–Southern Oscillation (ENSO) and abrupt Dansgaard–Oeschger events~\cite{Zebiak87,Dansgaard84,Broecker89} in the climate system.

Traditional modeling approaches, such as statistical time series analysis or physically based climate models, often face limitations in capturing the full spectrum of variability observed in real-world climate data. Deterministic approaches capture large-scale structure but overlook random fluctuations. On the other hand, while stochastic or statistical methods can capture variability, these methods are highly sensitive to noise~\cite{Majda09,Dijk13}. Recently, machine learning approaches using deep learning models, such as WaveNet-LSTM and AI-driven down-scaling frameworks, have demonstrated improved performance, such as in regional precipitation prediction~\cite{waqas2024,waqas2024seasonal,waqas2025}. 

Although significant advances have been made in understanding global climate variability, there remains a relative lack of emphasis on regional climate modeling, particularly in developing countries. Most global models are designed to capture broad climate patterns and may overlook the fine-scale spatial and temporal variability that governs regional climate systems. This limitation has led to a growing need for data-based, region-focused models that can help guide local climate planning and adaptation efforts. This research gap motivates our present study on the investigation of the climate variability, particularly maximum temperature dynamics, observed for a regional place called Imphal, the capital city of the state of Manipur, which is located in the northeastern part of India. This region, a part of the Indo-Burma biodiversity hotspot, experiences complex weather patterns influenced by orography, monsoon systems, and large-scale atmospheric variability~\cite{Jain13,Goswami06,Sreekesh16}. Despite its ecological and climatic significance, Northeast India remains understudied in the context of quantitative climate modeling.

In this work, we develop a hybrid dynamical-stochastic model of maximum temperature variability for Imphal, using publicly available data recorded over a time-span of 73 years. Our proposed data-driven hybrid model explicitly links three components: 1) the deterministic backbone of the temperature variability, which is obtained through harmonic and spectral decomposition, 2) the stochastic fluctuations arising due to intrinsic and external perturbations, and 3) a nonlinear feedback mechanism that incorporates chaotic features, which is often observed in empirical climate data~\cite{Berner17}. The model developed is particularly relevant for the climatic setting of Imphal. Topographically, Imphal city is situated in a valley enclosed by hill ranges in all four directions. It experiences all four seasons - winter, spring, summer with monsoon, and autumn, making it an ideal regional location for temperature variability analysis, as the region is subjected to variable climatic patterns at both local and large temporal scales. To the best of our knowledge, a comprehensive data-driven quantitative modeling of maximum temperature dynamics for Imphal is not yet available in the literature.

To evaluate our model's effectiveness, we perform statistical tests using the coefficient of determination ($R^2$)~\cite{Wright21,Ozer85}, root mean square error ($RMSE$)~\cite{Armstrong92} and Kling-Gupta Efficiency ($KGE$)~\cite{Gupta11}. Furthermore, we compare the complexities of observed and modeled temperature time series using the nonlinear analysis method of the complexity–entropy ($CH$)-causality plane~\cite{Rosso07,Zunino12}, ensuring that the hybrid model replicates the underlying dynamical structure. A key feature of our study is the explicit characterization of noise, from which we develop stochastic differential equations known as Langevin equations, that reflect both deterministic and random components~\cite{Watkins2024}. We further derive associated Fokker–Planck equations, providing a theoretical ground that links the model's stochastic dynamics and the probabilistic evolution of the temperature states.

The paper is organized as follows: Section~\ref{sec:methods} describes the data used in our analysis, the workflow pipeline, and the theoretical models. Section~\ref{results} presents the main findings and results of our analysis. We provide concluding remarks and implications of our study in Section~\ref{sec:conc}.

\section{Methods}
\label{sec:methods}
\begin{figure*}
   \centering
   \includegraphics[scale=0.6]{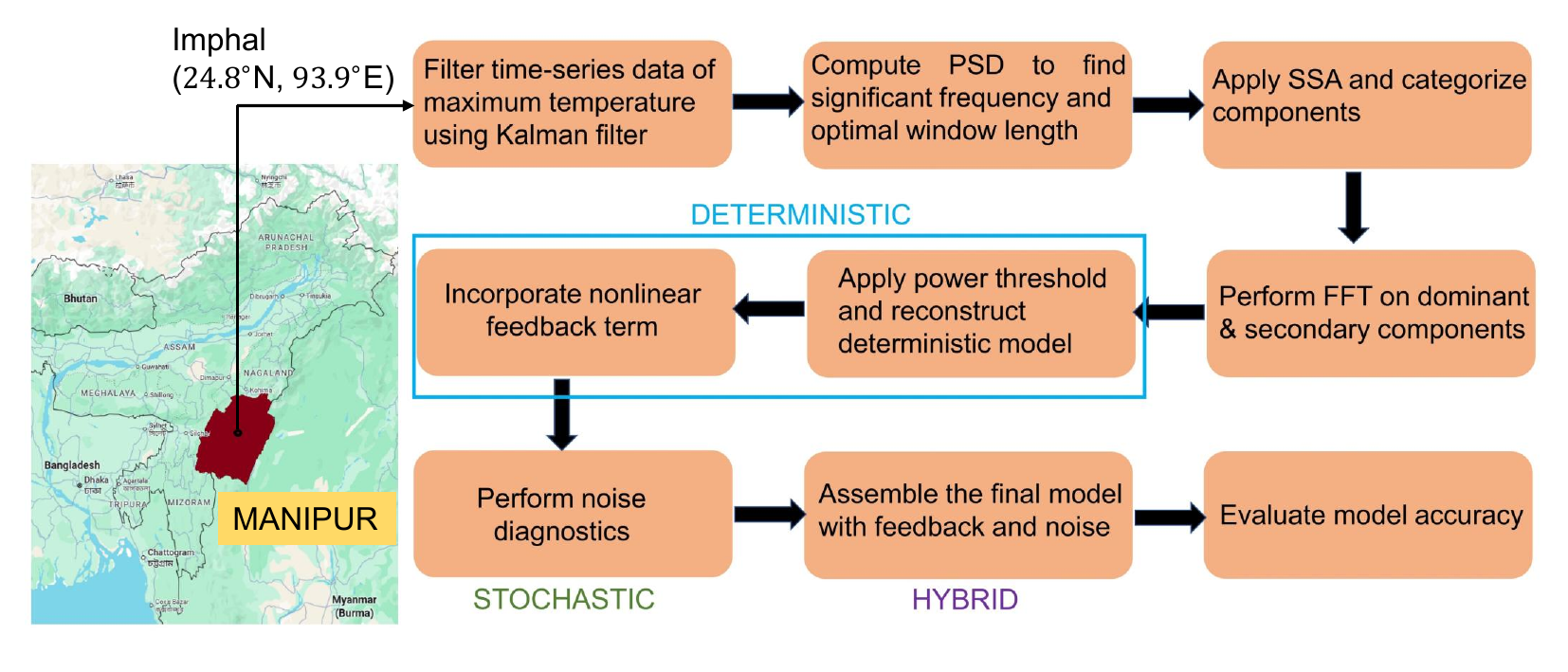}
   \vspace{-0.5cm}
   \caption{\textbf{Flowchart of our hybrid modeling framework:} The modeling begins with the filtering of the maximum temperature $T_{\textrm{max}}$ time series data and then applying singular spectrum analysis (SSA) after determining the optimal window length. Next, fast Fourier transform (FFT) is applied to the dominant and secondary components, after which we construct the deterministic part of the model. Noise diagnostics are then performed on the noise component. Lastly, we assemble the final model by adding the feedback and noise terms to the deterministic part.}
   \label{Fig1}
\end{figure*}
To analyze the variability of maximum temperature observed for Imphal, we develop a hybrid model that integrates three main components: (i) a \textit{deterministic} component, $\Lambda(t)$, derived from data decomposition using harmonic and spectral analysis, (ii) a \textit{nonlinear feedback} term, $F(t)$, accounting for temporal dependencies, and (iii) a \textit{stochastic} component, $\zeta(t)$, representing random fluctuations and characterized through noise diagnostics. We therefore formulate the hybrid model for maximum temperature, denoted by $\hat{T}_{\mathrm{max}}(t)$, as: 

\begin{eqnarray}
    \label{fin_model}
    \hat{T}_{\mathrm{max}}(t)=\Lambda(t)+F(t)+\Gamma\zeta(t), \label{eq:hm}
\end{eqnarray}
where $\Gamma$ is a dimensionless noise coefficient.

In the following subsections, we provide a detailed description of the construction of each component in Eq.~\eqref{fin_model}. Our framework follows the sequential data-processing and modeling pipeline as illustrated in Fig.~\ref{Fig1}. We have accessed the map indicating the geographic location details of Imphal from Google Maps~\cite{GoogleMapsImphal2025}.

\subsection{Data and Preprocessing}
For our present study, we utilize climate data from the Indian Meteorological Department (IMD) that provides a recent open-source Python library known as $\texttt{IMDlib}$~\cite{Nandi24,swain2024,tsela2025}. From this, we obtain data on daily maximum temperatures, denoted by $T_{\textrm{max}}$, from January 1951 to December 2024. We access the monthly data by dividing each annual dataset into 12 parts. We then proceed to perform our modeling approach for each month, thereby removing seasonal variations over the year. Outliers in the data (if present) are removed by imposing a condition to neglect $T_{\textrm{max}}<0 ^\circ C$ and $T_{\textrm{max}}>50 ^\circ C$, in accordance with realistic temperature records of Imphal.

To handle missing values and data gaps, we employ the Kalman filter~\cite{Kalman60,Welch95,Bellsky2014}, which simultaneously performs state estimation and data reconstruction. The Kalman filter provides an optimal recursive estimation framework that can reconstruct missing values based on system dynamics and observational uncertainty, as adopted by previous studies~\cite{moreno2020,haidu2025,fasso2023}(see Appendix~\ref{sec:kf} for details). We then decompose the filtered data using Singular Spectrum Analysis (SSA)~\cite{Hassani07,Golyan01,Kume16}, which initially estimates an optimal time window length for constructing a trajectory matrix by analyzing the Power Spectral Density (PSD) of the time series using the Welch method~\cite{Welch03}. From the frequency spectrum, we then identify the frequency with the maximum power as the dominant frequency, and hence its reciprocal is the optimal window length. We choose this PSD-based approach rather than cross-validation or sensitivity analysis because it yields a physically interpretable measure of temporal coherence. The PSD-based window selection links the window length directly to the intrinsic frequency structure of the signal, ensuring that the spectral decomposition captures the most relevant dynamical modes~\cite{SunLi2017,broomhead1986}.

Next, we convert the $T_{\textrm{max}}$ time series into the trajectory matrix using the Hankel transformation~\cite{Ghil02,Vautard89} that preserves the temporal relationships within the data. We then subject this matrix to Singular Value Decomposition (SVD)~\cite{Eckart36,Golub13}, where a threshold of 1\% of the maximum singular value from SVD is used to distinguish signal from noise. Based on this criterion, the SSA components are grouped into three categories: the dominant component (capturing the long-term trend), the secondary component (representing subtle or moderate oscillations), and the noise component (reflecting high-frequency random fluctuations). These dominant and secondary components together constitute the deterministic part of the $T_{\textrm{max}}$ variability, which we describe in the next subsection.

\subsection{Deterministic Component}
We construct the deterministic term $\Lambda(t)$ of our hybrid model~\eqref{eq:hm}, using the dominant and secondary components derived from SSA using Fast Fourier Transform (FFT)~\cite{Shen11,Saidi94,Trefethen00,de22} via a daily sampling interval. We thus express $\Lambda(t)$ as a truncated Fourier series~\cite{Stoica05}, retaining only the most influential periodic features of the dominant and secondary components:
\begin{eqnarray}
\label{deterministic}
    \Lambda(t)=\sum_i A_i^{(1)}\cos\left(2\pi f_i^{(1)}t+\phi_i^{(1)}\right)\nonumber\\
    +\sum_j A_j^{(2)}\cos\left(2\pi f_j^{(2)}t+\phi_j^{(2)}\right),
\end{eqnarray}
where $A$ denotes the amplitude (in  $^\circ $C units), $f$ represents the frequency and $\phi$ is the phase. The first summation is for the $i$ dominant components, and the second summation is for the $j$ secondary components. The number of terms in each summation depends on the power threshold chosen for each month. We ensure that both the mean and variance of the model output $\hat{T}_{\mathrm{max}}$ align with those of the original $T_{\mathrm {max}}$ time series.

\subsection{Feedback term}
Empirical observations indicate that feedback mechanisms significantly shape climate variability and transitions in real-world climate dynamics~\cite{Roe07,Bony06,Tsuchida2023,Dirmeyer2022}. To account for such nonlinear dynamics and temporal dependencies inherent in our $T_{\textrm{max}}$ time series data, we consider two types of feedback mechanisms for the feedback term $F(t)$ of our hybrid model~\eqref{eq:hm} as follows:
\begin{enumerate}
   \item \textit{Nonlinear Cubic feedback:}
    \begin{eqnarray}
    \label{cubic_fb}
    F(t)=\left\{
    \begin{array}{ll}
        0, & t<\Delta, \\
        \epsilon_1[T_{\mathrm{max}}(t-\Delta)]^3, & t\ge \Delta,
    \end{array}
    \right.
    \end{eqnarray}
    where $\Delta$ denotes the delay (lag) in days and $\epsilon_1$ is the feedback strength (in units of $^\circ$C$^{-2}$). Such a nonlinear cubic feedback is motivated by the fact that cubic nonlinearity captures regime switching that often occur in climate dynamics, such as the transition between stable and unstable temperature equilibria~\cite{Ghil94,Ashwin12,Thompson11}.  
\item \textit{Lorenz feedback:}
    \begin{eqnarray}
    \label{Lorenz_fb}
    F(t)=\left\{
    \begin{array}{ll}
        0, & t<\Delta, \\
\displaystyle\epsilon_2\frac{\sqrt{[y(t-\Delta)]^2+[z(t-\Delta)]^2}-\mu}{\sigma}, & t\ge \Delta, 
    \end{array}
    \right.
    \end{eqnarray}
    where $\Delta$ is the delay (in days), and $\epsilon_2$ is the feedback strength (in units of $^\circ$C). The parameters $\mu$ and $\sigma$ are respectively the mean and standard deviation of the Lorenz feedback signal that are derived from the $y(t)$ and $z(t)$ solutions of the Lorenz equations~\cite{Lorenz63}.
\end{enumerate}

We have considered the two feedback terms of~\eqref{cubic_fb} and~\eqref{Lorenz_fb} in order to examine their ability to reproduce the nonlinear variability inherent in the $T_{\textrm{max}}$ dynamics. We briefly explain the motivation behind incorporating such feedback types in the following. A cubic feedback term can introduce bistable or oscillatory behaviors depending on the sign of $\epsilon_1$, allowing the hybrid model~\eqref{eq:hm} to represent temperature-dependent processes such as radiative damping or surface–atmosphere coupling~\cite{Stommel61}. Moreover, cubic feedback terms generally appear in reduced climate models (e.g., energy balance or box models) through nonlinear expansions of radiative forcing and albedo response functions~\cite{Saltzman62}. Thus, a feedback type~\eqref{cubic_fb} can provide a minimal yet physically grounded way to emulate nonlinear regulation of temperature dynamics. On the other hand, a chaotic feedback is motivated by studies in nonlinear geophysical modeling, where chaotic dynamics are used to model internal variability or external chaotic forcing in climate systems~\cite{Dijk13,Ashwin12,Tel06}. In particular, a Lorenz feedback type of~\eqref{Lorenz_fb} can capture the influence of chaotic large-scale atmospheric and convective processes on regional temperature variability~\cite{Lorenz63,Saltzman62,Tsonis12}. By projecting the $y$ and $z$ components of the Lorenz system, which are related to convective temperature and vertical velocity~\cite{palmer1993}, onto a scalar feedback signal $F(t)$, the hybrid model~\eqref{eq:hm} incorporates external chaotic modulation that drives intermittent and irregular fluctuations in local temperature. Normalizing the signal using its mean $\mu$ and standard deviation $\sigma$ ensures that the feedback has a comparable amplitude to the observed $T_{\textrm{max}}$ variability while preserving the temporal complexity of the underlying chaotic dynamics. Such a Lorenz feedback further allows stochastic-like forcing of regional temperature dynamics by larger-scale circulation patterns, such as ENSO-like oscillations~\cite{Dijk13,Lucarini12}, thereby providing a physically grounded mechanism for capturing the irregular, non-periodic oscillatory components inherently present in $T_{\textrm{max}}$ dynamics.

\subsection{Noise diagnostics}
To determine the features of the noise component $\zeta(t)$ of our hybrid model~\eqref{eq:hm}, we initially perform noise diagnostics. The diagnostics include: (i) Kernel Density Estimation (KDE), (ii) Stability (denoted by $\alpha$) and Skewness parameter (SP) estimation, (iii) Power Spectral Density (PSD) and Spectral Decay parameter (denoted by $\beta$) estimation, and (iv) Additive/Multiplicative test. We refer the reader to Appendix~\ref{sec:nd} for their detailed descriptions. We briefly describe the key steps as follows: From KDE, we estimate the probability distribution of the noise component, and hence its mean and standard deviation. We determine the stability parameter ($\alpha$) and skewness parameter (SP) by fitting a Lévy alpha-stable distribution and estimate the thickness of the tail and symmetry of the distribution. From the PSD, we determine the spectral decay parameter ($\beta$), which specifies the characteristics of the noise. Further, the additive/multiplicative test identifies whether the noise component is dependent on the signal amplitude of the dominant and secondary components.

Depending on the results of our initial noise diagnostics (we will discuss in Section~\ref{results}), we consider three types of noise $\zeta(t)$:
\begin{itemize}
    \item \textit{White Noise} - %
    When $\zeta(t)$ is modeled as a standard Wiener process, its derivative $\displaystyle\frac{d\zeta(t)}{dt}$ corresponds to Gaussian white noise $\eta(t)$ with zero mean and delta-correlated fluctuations as $\langle\eta(t)\eta(t^\prime)\rangle=\delta(t-t^\prime)$. It is characterized by constant power spectral density across all frequencies. Gaussian white noise represents memoryless (Markovian) noise, which is used extensively in stochastic modeling~\cite{Gardiner85,van1992}. 
    \item \textit{Colored Noise} - Colored noise, characterized by frequency-dependent power spectra, has the autocorrelation function $\langle \eta(t)\eta(t^\prime)\rangle=\kappa(t-t^\prime)$ where $\kappa$ is not a delta function. Colored noise is often used to model systems with time-dependent fluctuations~\cite{west1990,rypdal2013}, for instance, the stochastic process of Ornstein-Uhlenbeck has $\kappa(t-t^\prime) \sim \displaystyle\frac{1}{2\tau} e^{-|t-t^\prime|/\tau}$ with $\tau$ as the correlation time of noise.
    \item \textit{Lévy noise} - Lévy noise arises from Lévy stable distributions $L(t)$, with the special case \small{$L(t)_{\alpha=1/2,\textrm{SP}=1,s_1,s_2}=\displaystyle\left(\frac{s_1}{2\pi}\right)^{1/2} \frac{1}{(t-s_2)^{3/2}}\exp\left[-\frac{s_1}{2(t-s_2)}\right], \ t>s_2$} \normalsize , where $\alpha$ is the stability parameter, SP denotes the skewness parameter, $s_1$ is the scale parameter and $s_2$ is the location (shift) parameter. Lévy noise generalizes Gaussian noise to include jumps or bursts, allowing for the modeling of anomalous diffusion stochastic processes, rare or extreme events, and systems with non-Gaussian fluctuations~\cite{sokolov2002,Metzler00,applebaum2009,metzler2014anomalous,LevyStable}.  
\end{itemize}

\begin{figure*}
    \centering
    \includegraphics[scale=0.5]{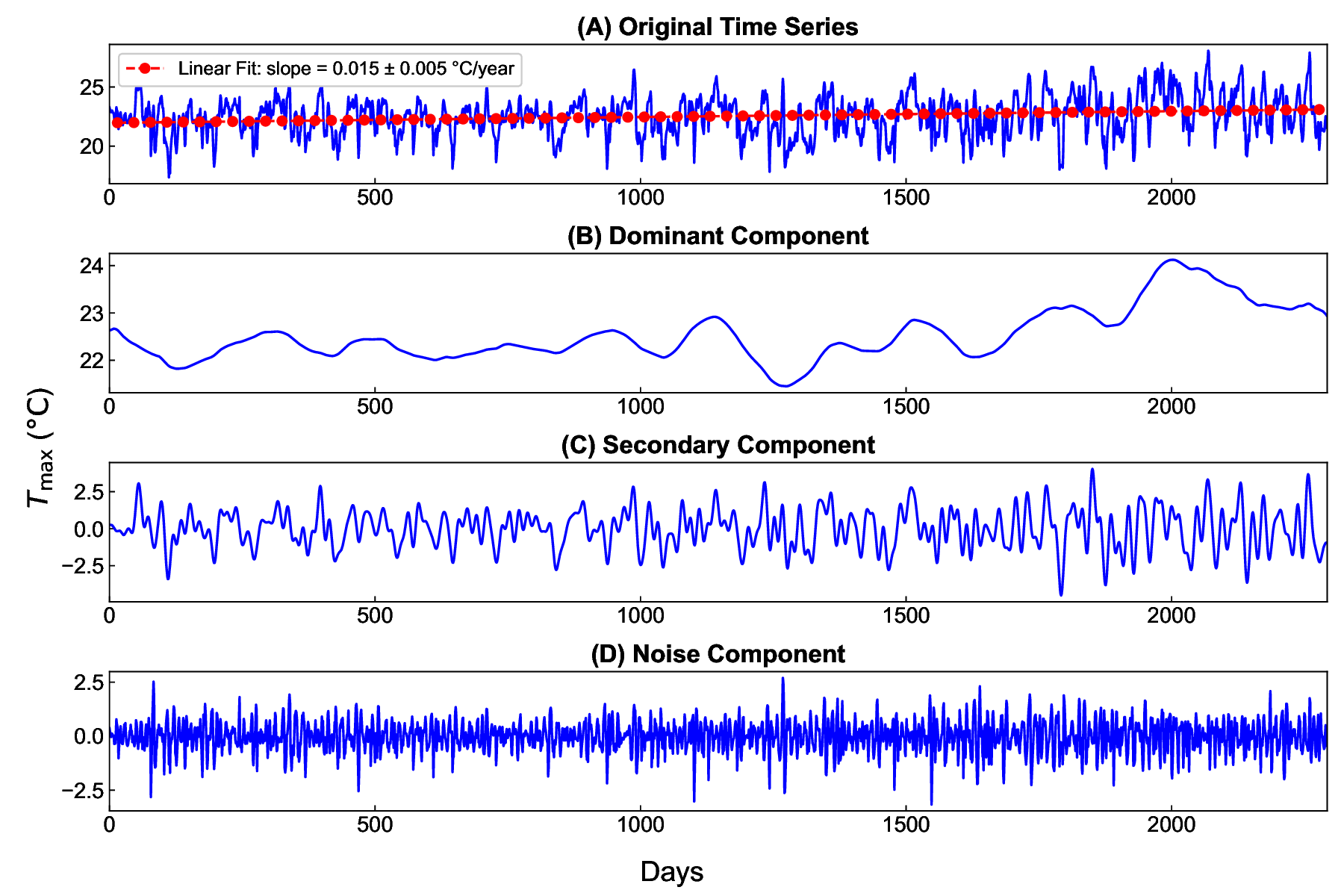}
    \caption{(A) The original time series of the monthly $T_{\mathrm{max}}$ data of Imphal for January for the period 1951-2024, obtained after Kalman filtering. The red dot dashed line indicates the linear regression fit to $T_{\mathrm{max}}$, where the slope indicates the gradual rise in the maximum temperature per year. (B) Dominant component, (C) Secondary component, and (D) noise component, obtained after applying singular spectrum analysis (SSA) to the $T_{\mathrm{max}}$ data.}
    \label{Fig2}
\end{figure*}

\section{Results \& Discussion}
\label{results}
We now present the results of the data-driven modeling of the observed maximum temperature $T_{\textrm{max}}$ data of Imphal using our hybrid model (Eq.~\eqref{fin_model}) on the time scale of months as well as the entire recorded time duration in years.

\subsection{SSA and Deterministic Model of monthly $T_{\textrm{max}}$ data}
Fig.~\ref{Fig2} presents the results of SSA applied to the monthly $T_{\textrm{max}}$ data of January from the year 1951 to 2024. In panel (A), we fit a linear regression line (red) to the original time series (blue) to estimate the increase in the maximum temperature of Imphal. From the slope of the regression line, we find that the maximum temperature $T_{\textrm{max}}$ of Imphal rises by $\sim0.015^\circ$C in January per year and leads to an increase of $\sim1.1^\circ$C in January over the period from 1951-2024. This increase of $T_{\textrm{max}}$ is a concerning indication of the significant warming trend in the region according to global climate change thresholds as identified by the Paris Agreement~\cite{UNFCCC2015} and the IPCC SR1.5 report~\cite{IPCC2018}. We also find similar warming trends for the other months (see the second column of Table \ref{Table1}). The largest cumulative increase in $T_{\textrm{max}}$ is observed in the month of November, aligning with known human perceptions of progressively warmer winters in Imphal over the years. Panels (B), (C), and (D) respectively show the dominant, secondary, and noise components for the $T_{\textrm{max}}$ time series of January, where the dominant and secondary components are seen to exhibit certain oscillatory behaviors.

\begin{figure}
    \centering
    \includegraphics[scale=0.34]{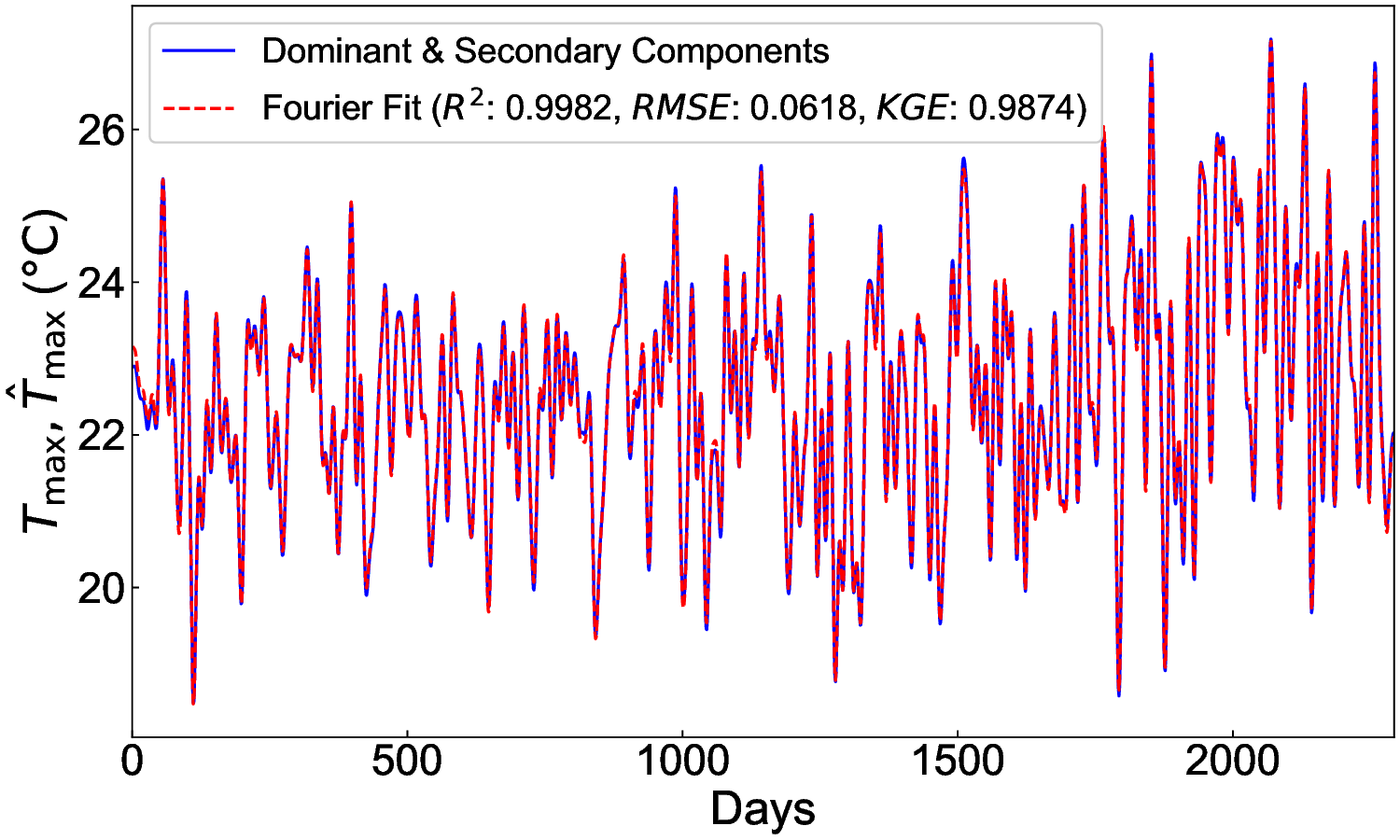}
    \caption{Blue curve indicates the dominant and secondary components extracted from the original $T_{\mathrm{max}}$ data. The red curve represents the deterministic model, as in the first term of $\hat{T}_{\mathrm{max}}$ in eq.~\eqref{fin_model}, obtained after performing the fast Fourier transform (FFT) on the dominant and secondary components.}
    \label{Fig3}
\end{figure}

\begin{table*}
\caption{\label{Table1}Statistical data for the different components of the hybrid model}
\begin{tabular}{|c|cc|cccccc|c|ccccc|}
\hline
\textbf{Month} & \multicolumn{2}{c|}{\textbf{Avg. Temp. Inc.}} & \multicolumn{6}{c|}{\textbf{Deterministic Model}} & \textbf{FB} & \multicolumn{5}{c|}{\textbf{Noise component}} \\
\cline{2-3} \cline{4-9} \cline{10-15}
 & per year & CI & DC & SC & $P_{\text{th}}$ & $R^2$ & $RMSE$ & $KGE$ & $\epsilon_2$ & M & SD & $\alpha$ & SP & $\beta$ \\
\hline
January & 0.015$\pm$0.005 \vline & 1.1$\pm$0.365 & 14 & 1146 & 0.1\% & 0.998 & 0.062 & 0.987 & 0.1 & 0 & 0.7 & 1.9 & -0.64 & -0.8 \\
February & 0.019$\pm$0.007 \vline & 1.4$\pm$0.511 & 40 & 611 & 0.1\% & 0.997 & 0.108 & 0.990 & 0.2 & 0 & 0.5 & 1.8 & -0.32 & -2.77 \\
March & 0.013$\pm$0.007 \vline & 1.0$\pm$0.511 & 39 & 692 & 0.1\% & 0.997 & 0.113 & 0.987 & 0.2 & 0 & 0.5 & 1.8 & -0.32 & -2.58 \\
April & 0.001$\pm$0.007 \vline & 0.1$\pm$0.511 & 9 & 1110 & 0.1\% & 0.997 & 0.097 & 0.987 & 0.2 & 0.01 & 1 & 1.9 & -0.57 & -0.66 \\
May & 0.011$\pm$0.005 \vline & 0.8$\pm$0.365 & 24 & 1146 & 0.1\% & 0.994 & 0.126 & 0.984 & 0.2 & 0 & 0.7 & 1.9 & -1 & -1.75 \\
June & 0.019$\pm$0.004 \vline & 1.4$\pm$0.292 & 16 & 1108 & 0.1\% & 0.997 & 0.088 & 0.988 & 0.2 & 0 & 0.7 & 1.9 & 0.14 & -1.42 \\
July & 0.020$\pm$0.004 \vline & 1.5$\pm$0.292 & 242 & 1145 & 0.005\% & 0.999 & 0.034 & 0.999 & 0.15 & 0 & 0.6 & 2 & -1 & -1.13 \\
August & 0.021$\pm$0.004 \vline & 1.6$\pm$0.292 & 75 & 1147 & 0.005\% & 0.999 & 0.012 & 0.999 & 0.05 & 0 & 0.8 & 1.9 & -1 & -0.02 \\
September & 0.021$\pm$0.004 \vline & 1.6$\pm$0.292 & 21 & 1110 & 0.1\% & 0.989 & 0.144 & 0.991 & 0.15 & 0 & 0.7 & 1.9 & -0.06 & -1.02 \\
October & 0.024$\pm$0.004 \vline & 1.8$\pm$0.292 & 29 & 888 & 0.1\% & 0.994 & 0.131 & 0.977 & 0.2 & 0 & 0.6 & 1.8 & -0.34 & -1.99 \\
November & 0.030$\pm$0.004 \vline & 2.2$\pm$0.292 & 32 & 457 & 0.1\% & 0.994 & 0.123 & 0.990 & 0.15 & 0 & 0.5 & 1.7 & -0.17 & -1.99 \\
December & 0.023$\pm$0.004 \vline & 1.7$\pm$0.292 & 34 & 402 & 0.1\% & 0.997 & 0.084 & 0.986 & 0.1 & 0 & 0.4 & 1.7 & -0.12 & -1.94 \\
\hline
\end{tabular}
\vspace{0.5em}
\begin{flushleft}
CI: Cumulative increase,
DC: No. of dominant components, 
SC: No. of secondary components, 
$P_{\text{th}}$: Power threshold, 
$R^2$: Coefficient of determination, 
\textit{RMSE}: Root Mean Square Error, 
\textit{KGE}: Kling-Gupta Efficiency, 
\textbf{FB}: Feedback term, 
$\epsilon_2$: Lorenz feedback strength, 
M: Mean, 
SD: Standard deviation, 
$\alpha$: Stability parameter, 
SP: Skewness parameter, 
$\beta$: Spectral decay parameter.
\end{flushleft}
\end{table*}

We now examine the Fourier components in the observed $T_{\textrm{max}}$ data and list the number of dominant components (DC) and secondary components (SC) after the FFT analysis in the third column of Table \ref{Table1}. For the month of April, the number of dominant components is found to be significantly lower compared to the other months. This possibly arises due to the underlying climatic transition period from mild temperature spring months to a greater temperature increase in the early summer month of April in many parts of the Indian subcontinent. During this transition period, temperature patterns tend to be more variable~\cite{jha2024}, resulting in energy being spread across many components, with no strong frequency and hence fewer dominant components. On the other hand, the month of July, corresponding to the peak monsoon season in most regions of India, has the highest number of dominant components. The temperature dynamics during monsoon are greatly influenced by cyclic and large-scale rainfall patterns~\cite{befort2016}, leading to more structural periodic components and hence greater dominant components.

We have employed SSA and FFT in our analysis because of the following reasons. SSA allows representing the observed temperature variability as a sum of a slowly varying trend, oscillatory components, and stochastic noise~\cite{Ghil02,Hassani07}. We note that SSA decomposition is valid under the assumption of weak stationarity within the analyzed time window. Further, the Fourier analysis allows representing the temperature signal as a superposition of harmonic oscillations with the assumption of stationary frequency content over the analyzed time span. Thus, SSA and FFT provide complementary information: while SSA determines temporally coherent structures , Fourier decomposition quantifies their spectral significance in frequency space, thereby ensuring a physically interpretable and reproducible decomposition of the observed $T_{\textrm{max}}$ dynamics~\cite{Golyan01}.

Fig.~\ref{Fig3} presents the deterministic component (eq.~\eqref{deterministic}) of the hybrid model of eq.~\eqref{fin_model}, with the blue curve indicating the dominant and secondary components extracted directly from the SSA components and the red curve representing the Fourier Fit to the original $T_{\textrm{max}}$ time series data of January after performing FFT on the dominant and secondary components. The two time series are seen to match almost perfectly. To assess this matching, we employ three different statistical metrics, namely, coefficient of determination ($R^2$), root mean square error ($RMSE$), and Kling-Gupta efficiency ($KGE$) (see Appendix~\ref{sec:at} for descriptions). For the Fourier fit of January data, values of $R^2$, $RMSE$, and $KGE$ are given in the legend. We also find good matches between the original and model-generated time series for the remaining months; month-wise statistical metrics are provided in the third column of Table~\ref{Table1}.

\subsection{Final Modeling of monthly $T_{\textrm{max}}$ data}
We proceed to perform noise diagnostics on the noise component (shown in panel D of Fig.~\ref{Fig2}) of $T_{\textrm{max}}$ dynamics for the months of January from 1951-2024. Appendix Fig.~\ref{FigB1} shows the probability density function (PDF) of this noise component, where we observe the PDF of the noise value to be slightly skewed with a small tail on the left. We fit the PDF with a Lévy alpha-stable distribution to determine the stability parameter ($\alpha$) and skewness parameter (SP). The Lévy alpha-stable distribution, which generalizes a Gaussian distribution (when $\alpha=2$), captures extreme events or abrupt shifts in temperature dynamics~\cite{Nolan97}. The SP controls the asymmetry of the PDF. Refer to the last column of Table~\ref{Table1} for the estimated values of $\alpha$ and SP for $T_{\textrm{max}}$ dynamics of January.

Further, we determine the spectral decay parameter ($\beta$) estimated from the PSD (see Appendix~\ref{sec:nd}) of the noise component of January for the period from 1951-2024; the PSD plot is shown in Appendix Fig.~\ref{FigB2}. We see that $\beta$ for the $T_{\textrm{max}}$ dynamics of January is found to have a negative value (see the last column of Table~\ref{Table1}), indicating that the overall power spectrum of the noise component increases with increasing frequency, which is a characteristic feature of colored-noise. The last column of Table~\ref{Table1} further presents all statistical information and results of the noise diagnostics for every month.

Next, we investigate for any significant correlation between the noise component and the signal components of $T_{\textrm{max}}$ for the months of January from 1951-2024. Appendix Fig.~\ref{FigB3} shows the scatter plot for the noise amplitude against the amplitude of the dominant and secondary components, with the red line indicating a linear regression. We see no significant correlation, and hence conclude that the noise is additive in nature.

In our hybrid model of Eq.~\eqref{fin_model}, in addition to the deterministic term $\Lambda(t)$, we also couple the cubic or Lorenz feedback (Eq.~\eqref{cubic_fb} or~\eqref{Lorenz_fb}) to the noise component $\zeta(t)$. To generate $\zeta(t)$, we use random number generators based on the statistical information obtained from the above diagnostic analysis of the noise component. We employ a colored noise generator characterized by $\beta$ (estimated from the PSD of the noise component) and a Lévy noise generator parametrized by $\alpha$ and SP (estimated from Lévy alpha-stable distribution PDF-fit). We prefer to use Lévy noise over fractional Gaussian noise (fGn) as it allows infinite variance when $\alpha < 2$, which is more suitable for climate modeling. Moreover, as our noise component is slightly skewed, fGn becomes a poor fit. The use of colored noise (abbreviated as CN) will reflect temporal correlations in the $T_{\textrm{max}}$ dynamics, while Lévy noise (LN) will incorporate the possibility of having extreme and rare events characterized by sudden jumps in the dynamics~\cite{lucarini2022}. For comparison, we also include white noise (WN).

Now, we employ two well-known measures of complexity, namely, permutation entropy~\cite{Bandt02} and statistical complexity measure~\cite{lopez1995} to investigate and compare the complexities between the original time series (hereafter abbreviated as OTS) $T_{\textrm{max}}$ and the time series $\hat{T}_{\textrm{max}}$ from the hybrid model  (Eq.~\eqref{fin_model}). Based on Shannon entropy, permutation entropy (denoted by $H$) quantifies the degree of disorder in a time series (see definition in Appendix~\ref{sec:at}). On the other hand, statistical complexity (denoted by $C$) measures the degree of organization in a time series by estimating how much the so-called ordinal probability distribution deviates from a uniform distribution (see definition in Appendix~\ref{sec:at}). In particular, we use the two-dimensional representation known as complexity-entropy ($CH$) causality plane~\cite{Rosso07,Zunino12,pessa2021}, which captures not only disorder but also the degree of correlational structure in a given time series. The $CH$-plane is proven to be a powerful diagnostic tool for distinguishing dynamical processes of different physical origins, such as deterministic chaos, stochastic noise, or periodic signals~\cite{Rosso07}. It offers a model-free, data-driven approach to quantify complexity in any empirical time series data~\cite{chanu2024exploring}. Plotting the $(H,C)$ values of the OTS and the hybrid model with different feedback and noise types on the $CH$ plane allows us to evaluate how well the given hybrid model matches the complexity of the OTS, thereby validating our data-driven modeling results. 

\begin{figure}
    \centering
    \includegraphics[scale=0.3]{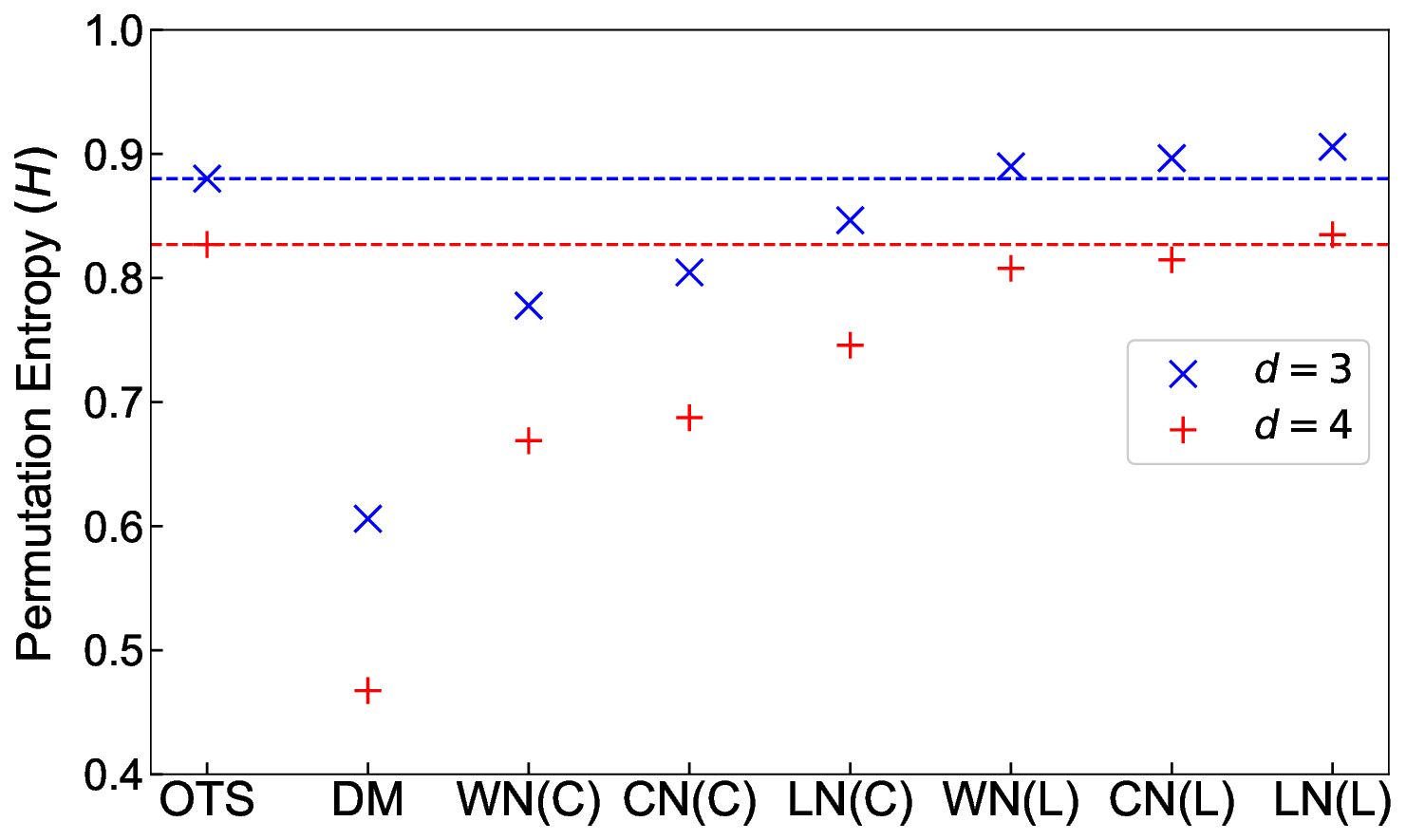}
    \caption{We compare the values of permutation entropy $H$ (using embedding dimension $d=3,4$) for the original time series (OTS) with deterministic model (DM), hybrid models incorporating different noise and feedback types (see $x$-axis labels). WN(C): White noise (Cubic), CN(C): Colored noise (Cubic), LN(C): Lévy noise (Cubic), WN(L): White noise (Lorenz), CN(L): Colored noise (Lorenz), LN(L): Lévy noise (Lorenz). The dotted blue and red lines indicate $H$ values for the OTS when $d=3$ and 4, respectively.}
    \label{Fig4}
\end{figure}

To evaluate and compare the degree of disorder, we first compute the value of $H$ for the $T_{\textrm{max}}$ OTS, deterministic term or model (DM) and six other possible models that incorporate combinations of different feedback and noise types in $\hat{T}_{\textrm{max}}$ (abbreviated as: WN(C): White noise (Cubic), CN(C): Colored noise (Cubic), LN(C): Lévy noise (Cubic), WN(L): White noise (Lorenz), CN(L): Colored noise (Lorenz), and LN(L): Lévy noise (Lorenz)). For the respective time series, we calculate $H$ using two different values of embedding dimension $d=3$ (blue crosses) and 4 (red plus markers), results shown in Fig.~\ref{Fig4}. Dashed blue and red lines indicate the $H$ values for the OTS. Fig.~\ref{Fig4} shows that for both embedding dimensions, the calculated values of $H$ for the three models of WN(L), CN(L), and LN(L) are close to that of the OTS, whereas that of DM is far from the OTS. These results show that DM fails to represent the real $T_{\textrm{max}}$ dynamics, whereas WN(L), CN(L), and LN(L) well capture the amount of disorder or uncertainty in the maximum temperature dynamics. 

\begin{figure}
    \centering
    \includegraphics[scale=0.3]{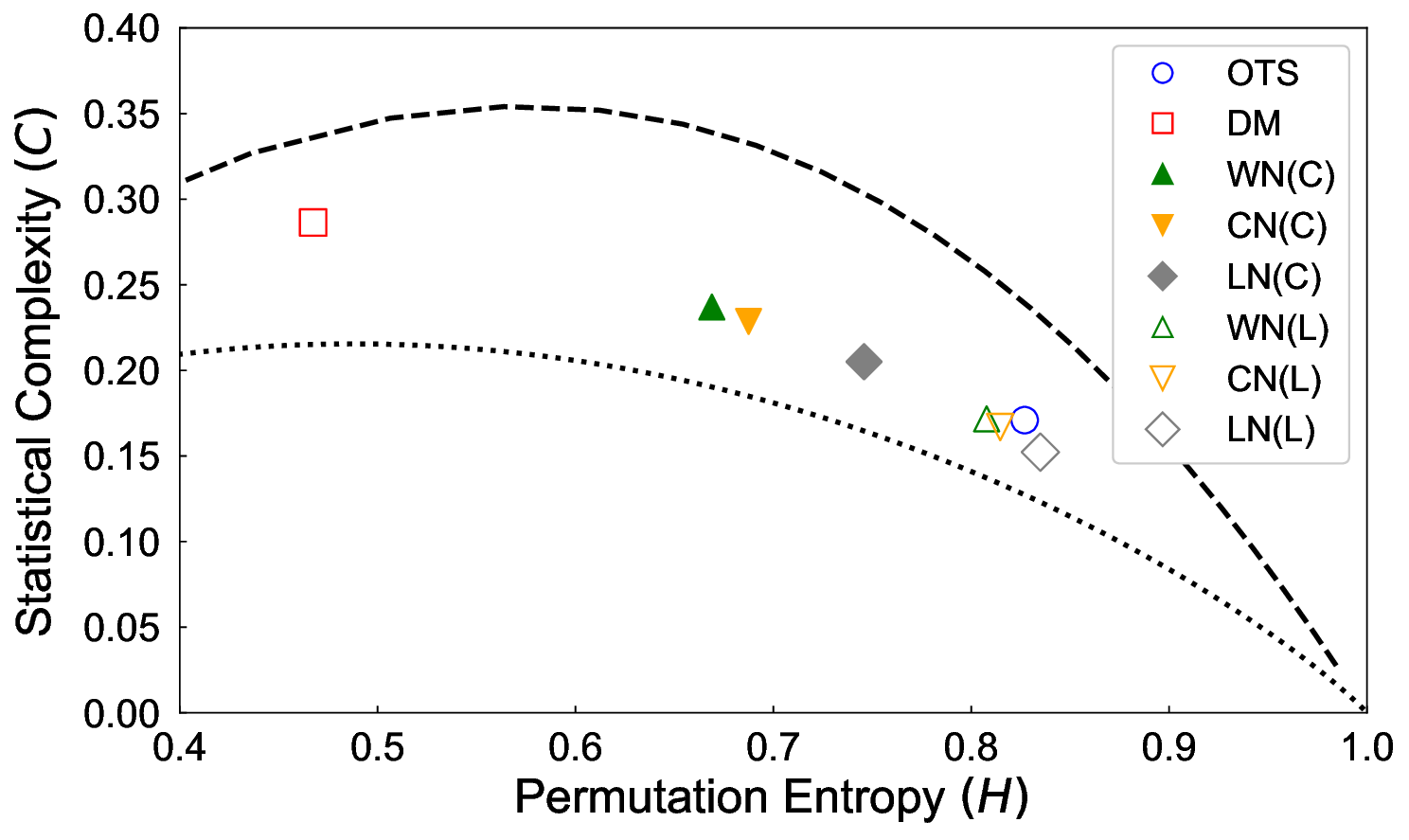}
    \caption{Complexity-Entropy ($CH$)-plane showing the values of ($H,C$) for the original time series (OTS), the deterministic model (DM), and the hybrid models for different noise and feedback configurations, for the monthly data of January of the period 1951-2024 (using $d=4$). In the legend, WN(C): White noise (Cubic), CN(C): Colored noise (Cubic), LN(C): Lévy noise (Cubic), WN(L): White noise (Lorenz), CN(L): Colored noise (Lorenz), LN(L): Lévy noise (Lorenz).}
    \label{Fig5}
\end{figure}

In addition to $H$, we also calculate the values of statistical complexity measure $C$ (using $d=4$), and plot the $(H,C)$ values of OTS, DM, WN(C), CN(C), LN(C), WN(L), CN(L), and LN(L) on the $CH$-plane of Fig.~\ref{Fig5} (see legend for marker types and colors). We see that the $(H,C)$ values of WN(L), CN(L), and LN(L) closely match those of the OTS as compared to DM, WN(C), CN(C), and LN(C). These results further support the result that Lorenz feedback models with white, colored, and Lévy noise effectively represent the real $T_{\textrm{max}}$ dynamics. In the plot, the dashed and dotted black curves indicate the theoretical maximum and minimum values of statistical complexity at each value of permutation entropy at a given embedding dimension~\cite{Martin06}.

\begin{figure}
   \centering
   \includegraphics[scale=0.34]{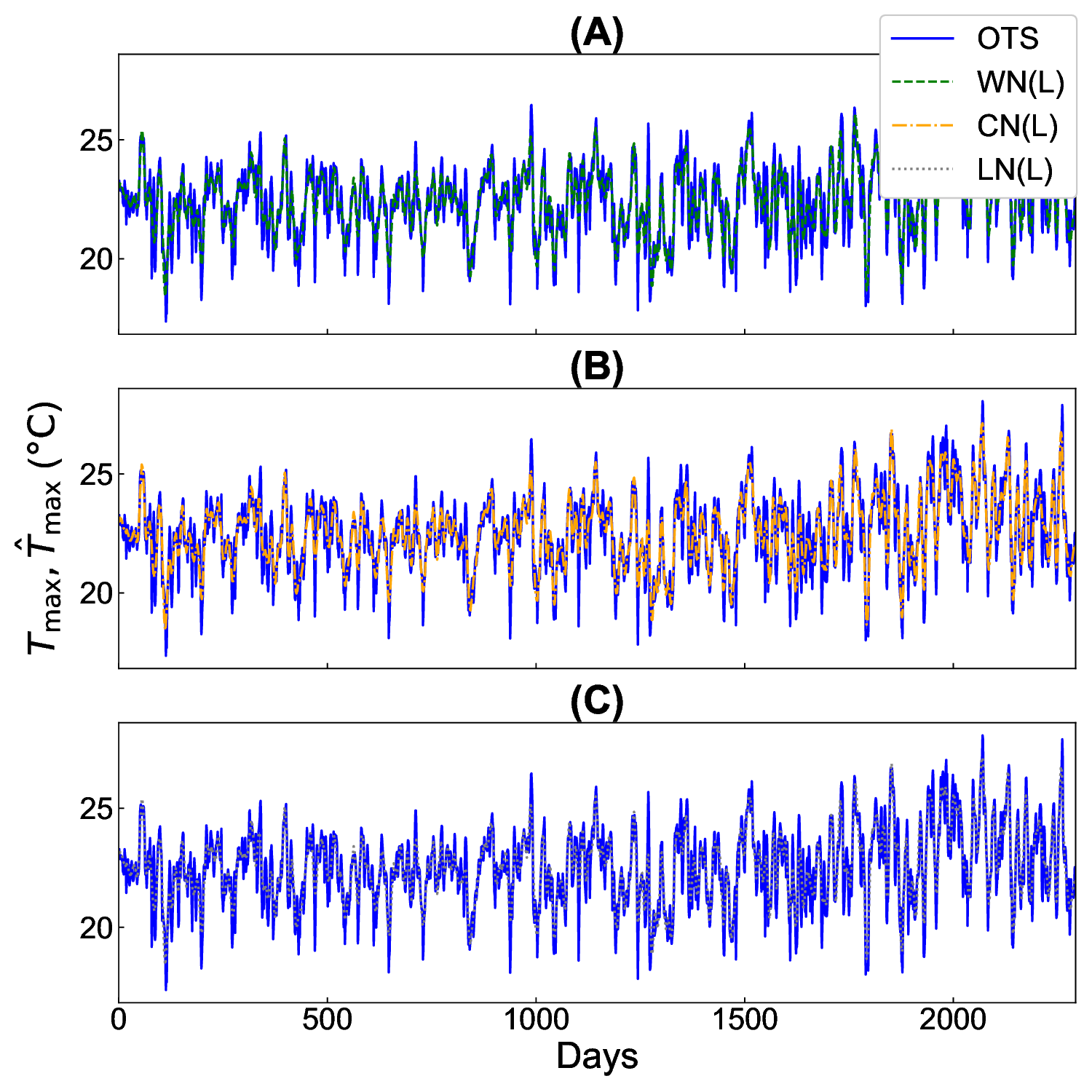}
   \caption{Plots of maximum temperature time series for the original data ($T_{\textrm{max}}$) and the hybrid model with Lorenz feedback term ($\hat{T}_{\textrm{max}}$) using noise types: (A) White noise (WN), (B) Colored noise (CN), and (C) Lévy noise (LN), for the monthly data of January of the period 1951-2024. In each subplot, blue curves indicate the original time series (OTS) of $T_{\textrm{max}}$.}
   \label{Fig6}
\end{figure}

To further analyze the observed good representation of real $T_{\textrm{max}}$ dynamics by Lorenz feedback models with white, colored, and Lévy noise, we now plot the simulated $\hat{T}_{\textrm{max}}$ time series of the WN(L), CN(L), and LN(L) models (Eq.~\eqref{fin_model}) along with the OTS in each panel of Fig.~\ref{Fig6}. Visual inspection reveals that the time series from the three models matches well with the original time series. The parameters used for simulation are: $\epsilon_1=10^{-6}$ $^\circ$C$^{-2}$ (for cubic feedback of Eq.~\eqref{cubic_fb}) and $\epsilon_2=10^{-1}$ $^\circ$C (for Lorenz feedback of Eq.~\eqref{Lorenz_fb}). In both feedback types, we use $\Delta=1$ day. For the noise coefficient in Eq.~\eqref{fin_model}, we consider $\Gamma=10^{-1}$. By embedding these physically motivated feedback mechanisms within a stochastic modeling framework, the hybrid model of eq.~\eqref{fin_model} provides a link between statistical representations of variability and underlying climatic processes, thereby offering both quantitative analysis and physical interpretability.

\begin{figure}
   \centering
   \includegraphics[scale=0.34]{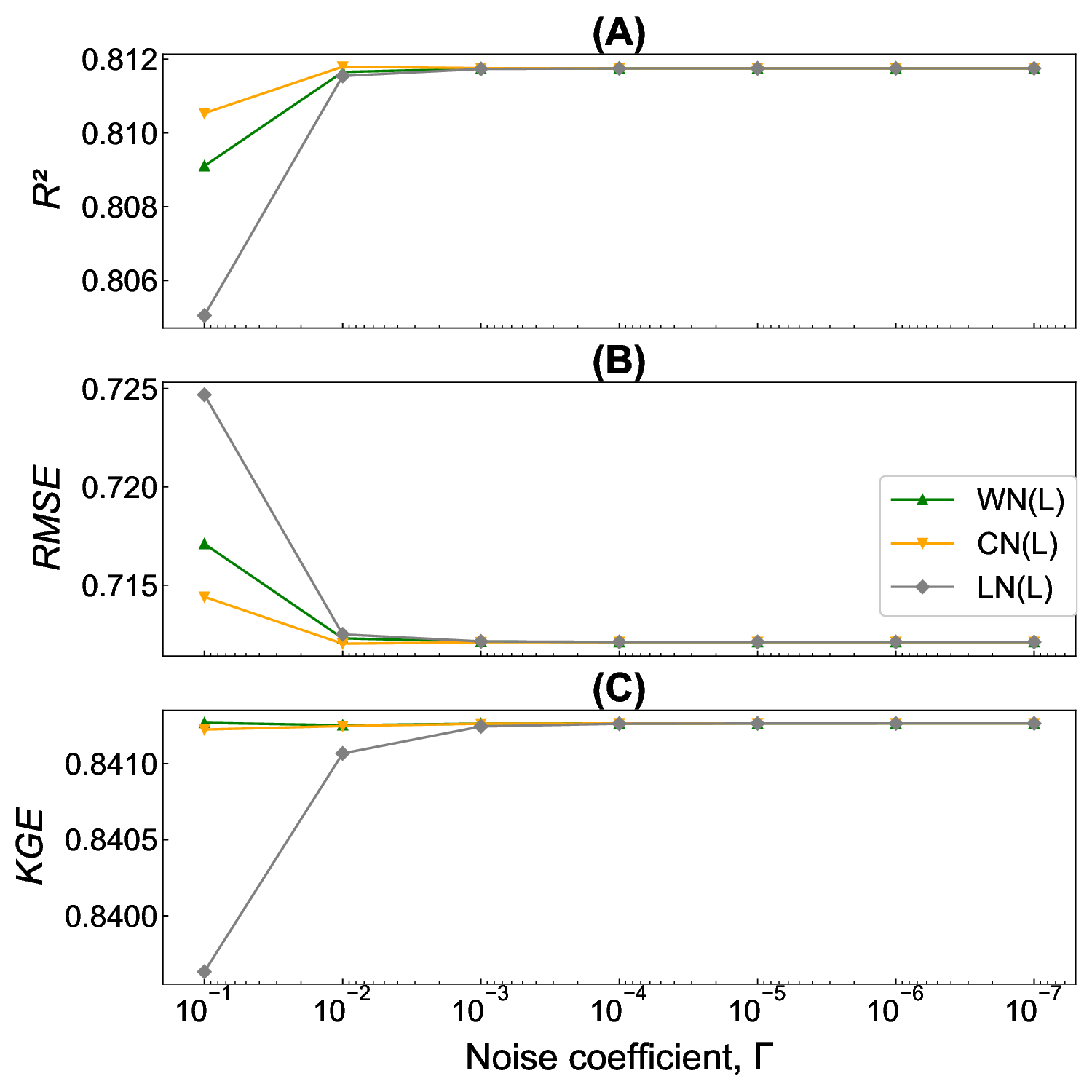}
   \caption{\textbf{Statistical accuracy tests of the hybrid model (with Lorenz feedback and different noise types) against the original time series:} We plot the variation of three statistical metrics (A) $R^2$, (B) $RMSE$, and (C) $KGE$ with noise coefficient ($\Gamma$) for the monthly data of January of the period 1951-2024 In the legend, WN(L): White noise (Lorenz), CN(L): Colored noise (Lorenz), LN(L): Lévy noise (Lorenz).}
   \label{Fig7}
\end{figure}

We now test the effect of noise strength, quantified by $\Gamma$, on the modeling accuracy of the simulated $\hat{T}_{\textrm{max}}$ from the models of WN(L), CN(L), and LN(L) with respect to the original time series. For this accuracy test, we use three different statistical metrics, namely, $R^2, RMSE$, and $KGE$. The results for varying $\Gamma$ are shown in Fig.~\ref{Fig7}. These results facilitate further testing of the robustness of our hybrid modeling approach when different strengths of stochastic forcing appear. We see that the values of the three metrics approach a fixed value when $\Gamma\le 10^{-3}$, indicating that the noise magnitude in the simulated time series has matched the level of stochasticity present in the observed $T_{\textrm{max}}$ data. The saturation values are $R_s^2\approx0.81$, $RMSE_s\approx0.71$, and $KGE_s\approx0.84$. Analysis, as shown in Fig.~\ref{Fig7}, also helps in tuning the model parameters to reproduce the observed variability present in the empirical data.

\begin{figure*}
    \centering
    \includegraphics[scale=0.35]{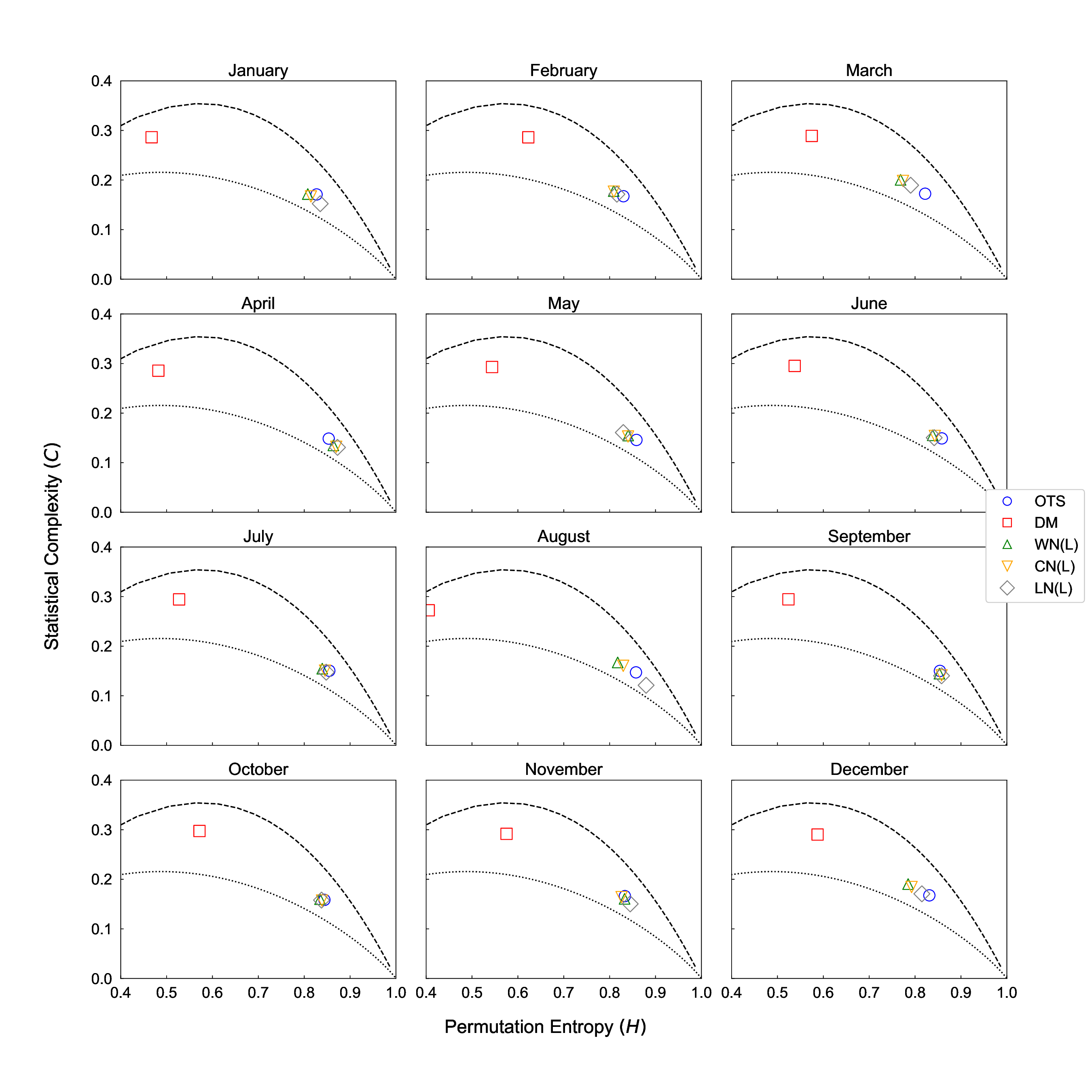}
    \vspace{-1cm}
    \caption{Complexity-Entropy ($CH$)-planes for each month (January to December) of the period 1951-2024 (using $d=4$). In the legend, OTS: Original time series, DM: Deterministic model, WN(L): White noise (Lorenz), CN(L): Colored noise (Lorenz), LN(L): Lévy noise (Lorenz).}
    \label{Fig8}
\end{figure*}

Additionally, Fig.~\ref{Fig8} presents the individual \textit{CH}-planes for all the months of the period from 1951-2024, showing the $(H,C)$ values of OTS, DM, WN(L), CN(L), and LN(L). Location of all the $(H,C)$ values of the OTS and the Lorenz feedback models of WN(L), CN(L), and LN(L) in the right side of the $CH$-plane with $H>0.5$ indicate that the observed dynamics of maximum temperature of Imphal is predominantly stochastic and is less likely to be chaotic.

We note that the $CH$-plane analysis depends on two parameter choices, namely, the embedding dimension ($d$) and the number of data points (or data length) ($N$). A very small value of $d$ may fail to capture the underlying dynamical structure accurately, whereas a very large $d$ not only results in high computational cost but can also introduce bias in probability estimation due to insufficient ordinal patterns when $d!\approx$$N$~\cite{Bandt02}. Moreover, short time series or those with strong seasonality can lead to statistical under-sampling of patterns, thereby affecting both $H$ and $C$ values~\cite{Zunino12}. To mitigate these issues, we have followed the commonly accepted criterion $d!\ll N$~\cite{riedl2013practical} and have adopted $d=3$ and $4$ to get a reliable statistical estimate of ordinal probability distribution for the monthly $T_{\textrm{max}}$ data used here.


\begin{figure*}
\centering\includegraphics[height=10.0cm,width=14.0cm]{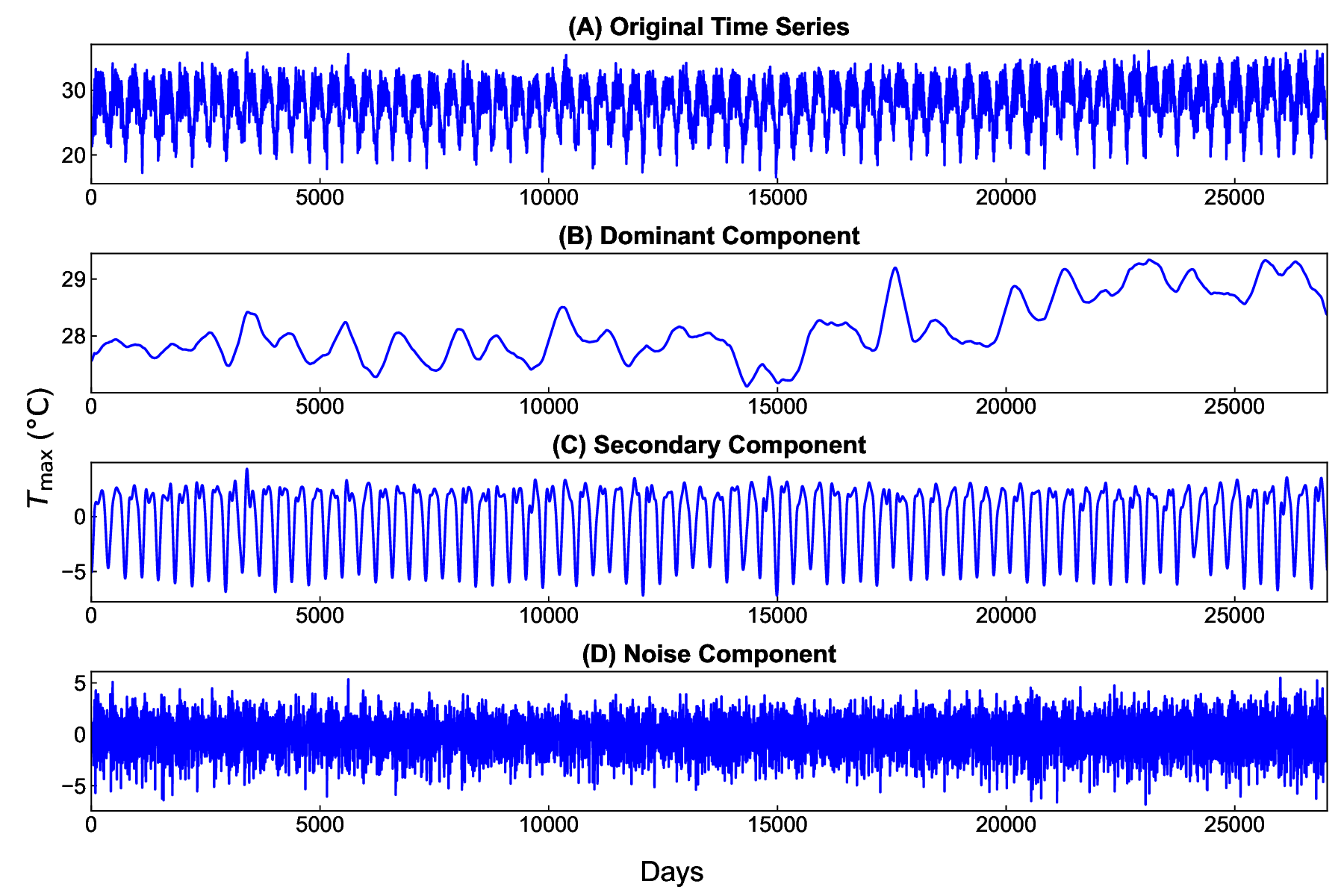}
    \caption{(A) The original time series of the maximum temperature $T_{\textrm{max}}$ data for the period 1951-2024 (obtained after Kalman filtering) without splitting into monthly data. (B) Dominant component, (C) secondary component, and (D) noise component are obtained after applying Singular Spectrum Analysis (SSA) to the $T_{\textrm{max}}$ data. }
    \label{Fig9}
\end{figure*}

\begin{figure}
    \centering
    \includegraphics[height=6.0cm,width=9.0cm]{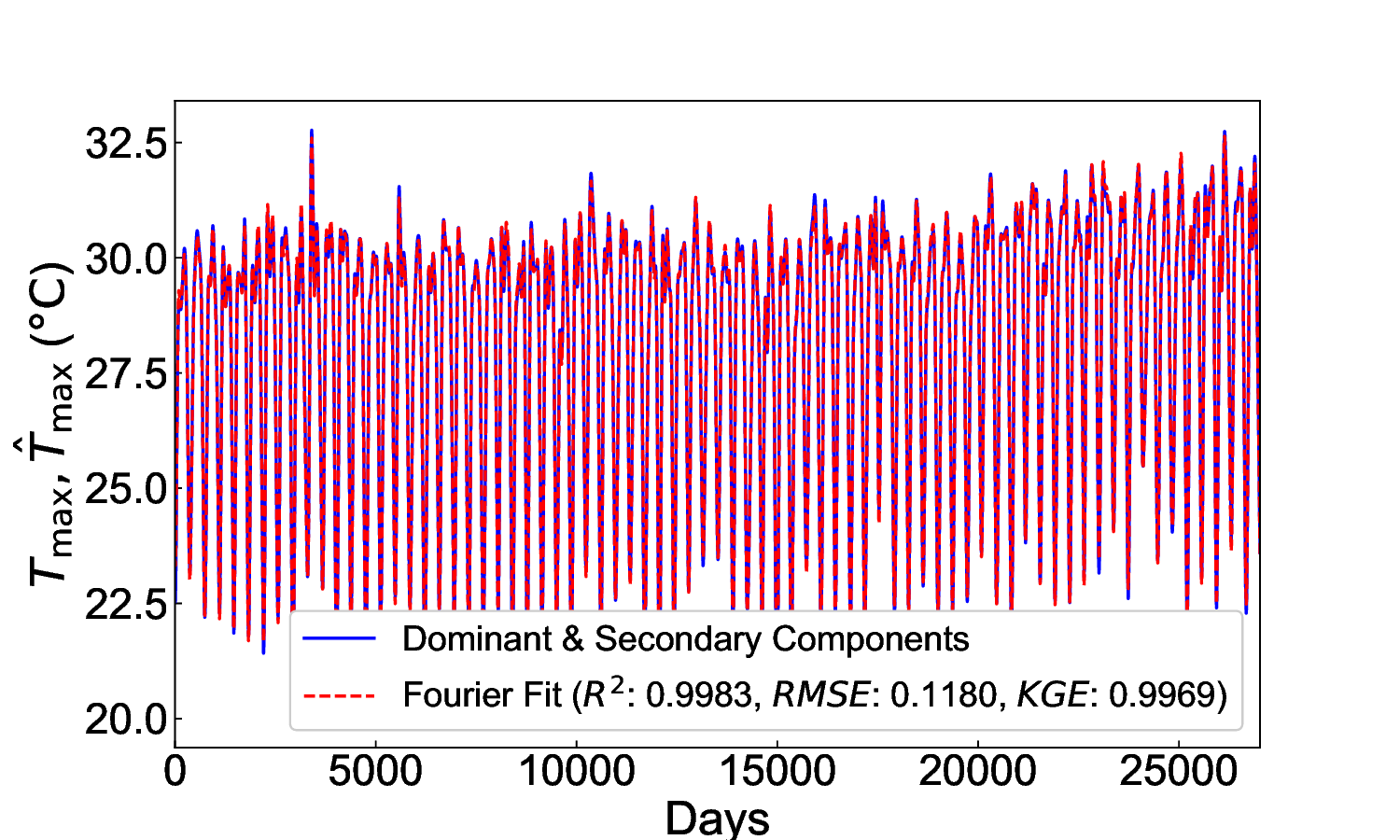}
    \caption{Blue curve indicates the dominant and secondary components extracted from the original maximum temperature time series data $T_{\mathrm{max}}$ for the entire time scale of the period 1951-2024, after applying SSA. The red curve represents the deterministic model, as in the first term of $\hat{T}_{\mathrm{max}}$ in eq.~\eqref{fin_model}, obtained after performing the fast Fourier transform (FFT) on the dominant and secondary components.}
    \label{Fig10}
\end{figure}
\begin{figure}
    \centering
    \includegraphics[height=6.0cm,width=8.0cm]{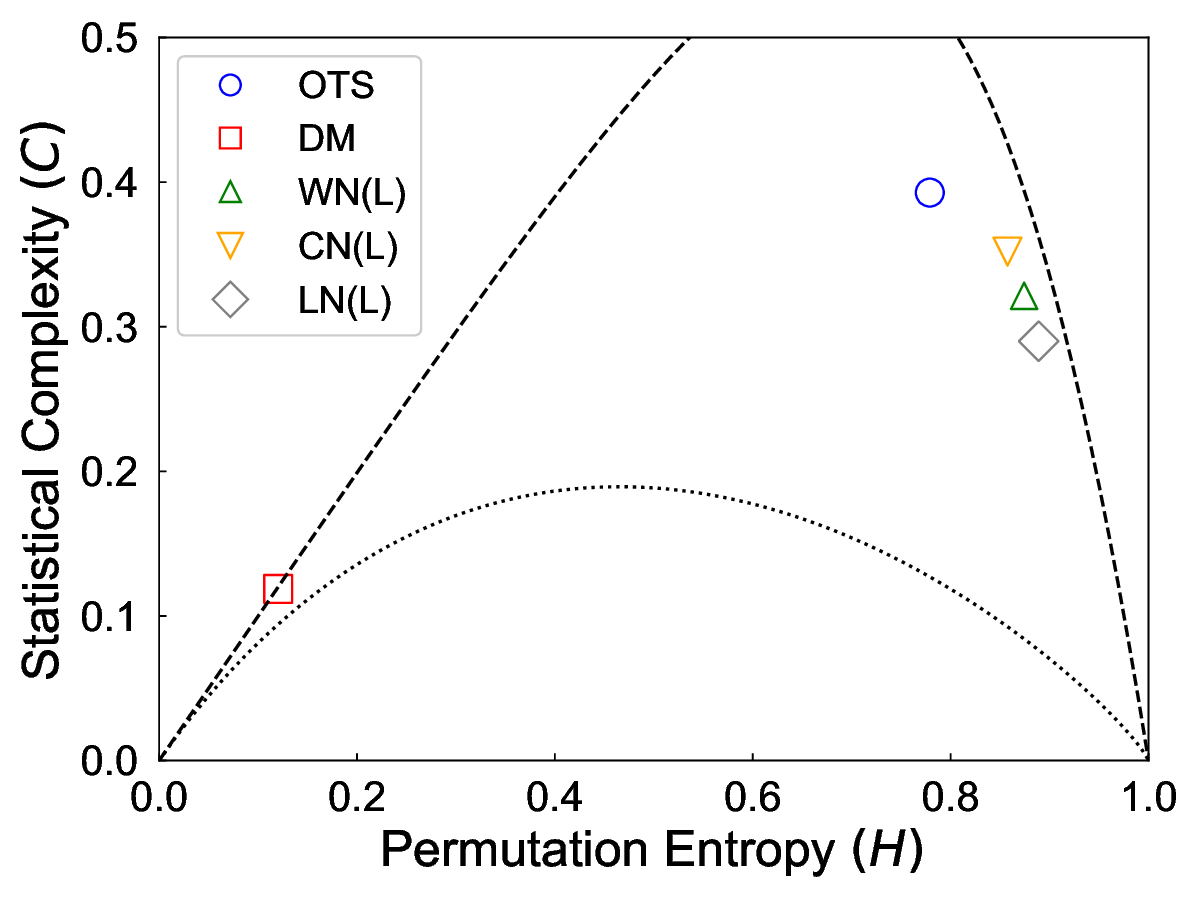}
    \caption{Complexity-Entropy ($CH$)-plane showing the values of ($H,C$) for the original time series (OTS), the deterministic model (DM), and the hybrid model with Lorenz feedback and different noise types, for the entire time scale data without splitting into monthly segments (using $d=7$). In the legend, WN(L): White noise (Lorenz), CN(L): Colored noise (Lorenz), LN(L): Lévy noise (Lorenz).}
    \label{Fig11}
\end{figure}
\begin{figure}
   \includegraphics[scale=0.33]{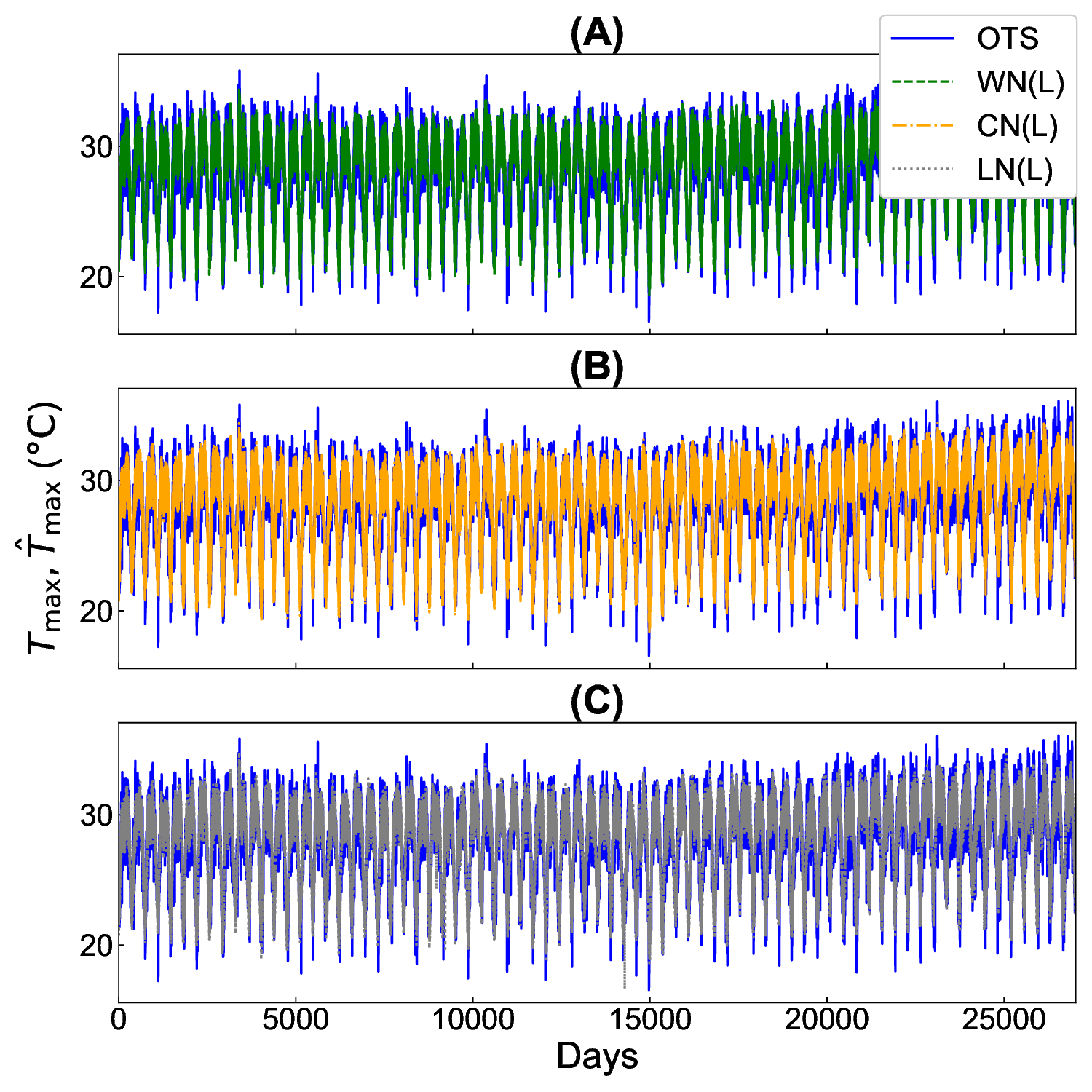}
   \caption{Time series plots of the maximum temperature original data ($T_{\textrm{max}}$) and the hybrid model with Lorenz feedback term ($\hat{T}_{\textrm{max}}$) using noise types: (A) White noise (WN), (B) Colored noise (CN), and (C) Lévy noise (LN), for the entire time scale data without splitting into monthly segments (using $d=7$). In each subplot, blue curves indicate the original time series (OTS) of $T_{\textrm{max}}$.}
   \label{Fig12}
\end{figure}

\begin{figure}
   \includegraphics[scale=0.33]{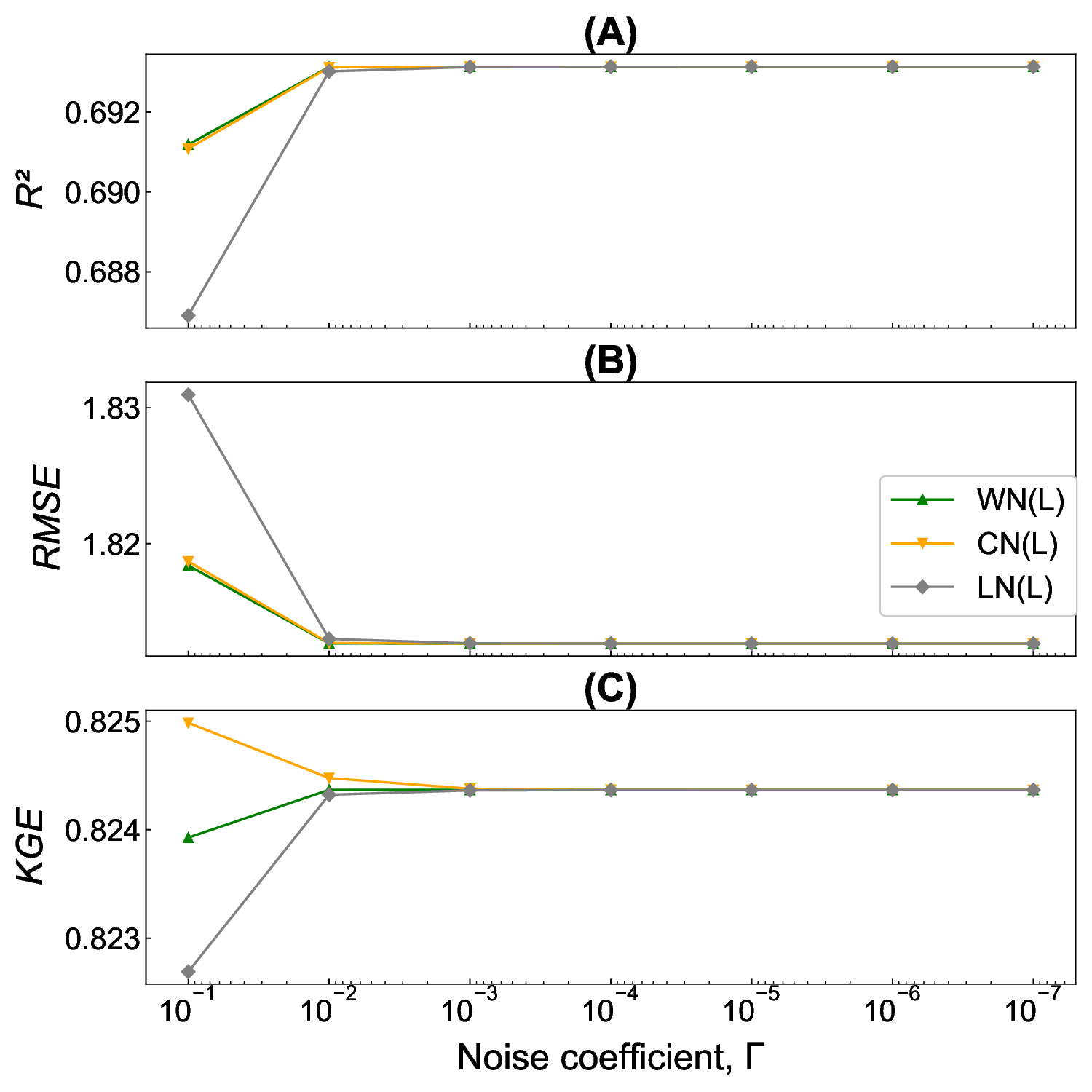}
   \caption{\textbf{Statistical accuracy tests of the hybrid model (with Lorenz feedback and different noise types) against the original time series:} We plot the variation of three statistical metrics (A) $R^2$, (B) $RMSE$, and (C) $KGE$ with noise coefficient ($\Gamma$) for the entire time scale data without splitting into monthly segments. In the legend, WN(L): White noise (Lorenz), CN(L): Colored noise (Lorenz), LN(L): Lévy noise (Lorenz).}
   \label{Fig13}
\end{figure}

\begin{figure}
    \centering
    \includegraphics[scale=0.25]{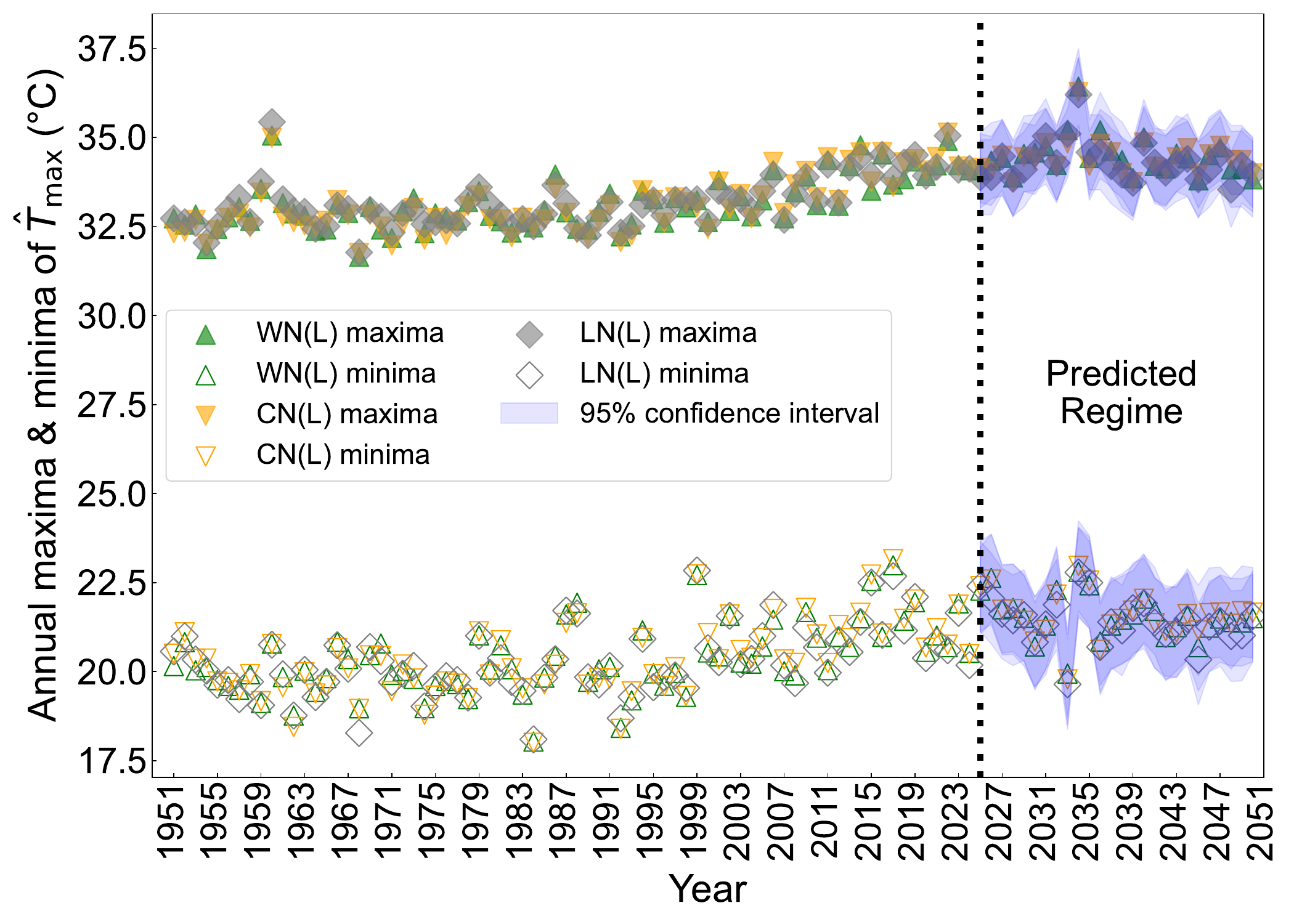}
    \caption{The annual maxima and minima of $\hat{T}_{\textrm{max}}$ time series of the hybrid model in Fig.~\ref{Fig12}. The model is extrapolated with a 95\% confidence interval to show the increasing trends of $\hat{T}_{\textrm{max}}$. The vertical dotted line indicates the current year of 2025. In the legend, WN(L): White noise (Lorenz), CN(L): Colored noise (Lorenz), LN(L): Lévy noise (Lorenz).}
    \label{Fig14}
\end{figure}

\begin{figure}
    \centering
    \includegraphics[scale=0.25]{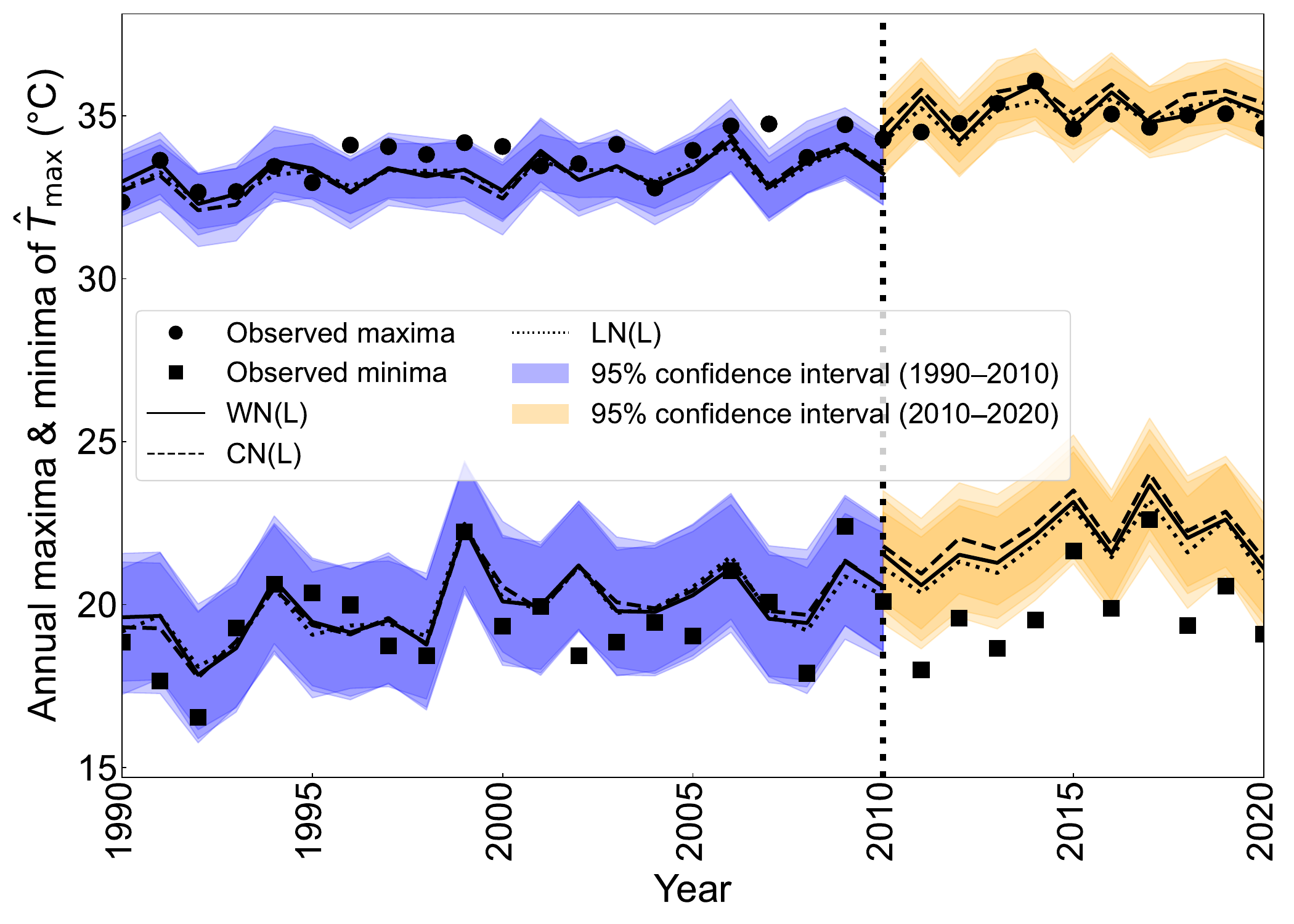}
    \caption{\textbf{Prediction skill assessment of the hybrid model using hindcast validation:} We perform the prediction skill assessment of the annual temperature extremes of the $\hat{T}_{\textrm{max}}$ predicted by the hybrid model, for the periods 1990-2010 and 2010-2020, corresponding to the result shown in the above Fig.~\ref{Fig14}. Shaded regions indicate 95\% confidence intervals. The vertical dotted line indicates the boundary between the two hindcast periods. In the legend, WN(L): White noise (Lorenz), CN(L): Colored noise (Lorenz), LN(L): Lévy noise (Lorenz).}
    \label{Fig15}
\end{figure}

\subsection{Modeling of $T_{\textrm{max}}$ data from January 1951 to December 2024}
We now repeat our analysis on the entire time series data of $T_{\textrm{max}}$ for the period from January 1951 to December 2024, without separating it into individual months; results presented in Figs.~\ref{Fig9} to~\ref{Fig13}. Fig.~\ref{Fig9}(A) presents the original $T_{\textrm{max}}$ data for the entire duration after Kalman filtering. Figs.~\ref{Fig9}(B), (C), and (D) show the dominant, secondary, and noise components, respectively. Fig.~\ref{Fig10} displays the deterministic model along with the Fourier fit (statistical metrics indicated in the legend). Fig.~\ref{Fig11} shows the $CH$-plane for the entire $T_{\textrm{max}}$ time series (OTS) as well as simulated time series from DM, WN(L), CN(L), and LN(L) using $d=7$. The three Lorenz models with different noise types exhibit distinct values of $H$ and $C$ compared to those in Fig.~\ref{Fig5}, suggesting different degrees of entropy or disorder and complexity or correlational structure in large-scale dynamics of $\hat{T}_{\textrm{max}}$. Fig.~\ref{Fig11} shows that the Lorenz model with colored noise produces a $(H,C)$ result closer to the real dynamics of OTS as compared to that of Lévy or white noise. We plot the simulated time series $\hat{T}_{\textrm{max}}$ from these three Lorenz models along with the original time series $T_{\textrm{max}}$ in Fig.~\ref{Fig12}. We perform the noise diagnostics and generate the noise terms using the statistical information and parameters obtained. To optimize the statistical metrics, we have used the model parameters $\epsilon_2=1$$^\circ$C and $\Gamma=10^{-1}$. Fig.~\ref{Fig13} shows that at a much larger timescale, when using the entire dataset length, the statistical accuracies of the models decrease. The statistical accuracy test results stabilize at $R^2_s\approx 0.69$, $RMSE_s\approx 1.8$, and $KGE_s\approx 0.82$. While not excellent, $R^2_s\approx 0.69$ is regarded as a good result in the case of climate systems that are inherently noisy and nonlinear. In climate modeling, $RMSE_s\approx 1.8$ is generally acceptable for daily or monthly temperature data~\cite{moriasi2007}. The value of $KGE_s\approx 0.82$ again suggests that the models capture the shape and distribution of the observed data well~\cite{knoben2020,moriasi2007}. Nonetheless, the observed overall decrease in statistical accuracy of the model for the large-scale dynamics over the entire time scale indicates the need to incorporate higher-order factors in our hybrid model~\eqref{eq:hm}. This is also evident from the previous result of $CH$-plane (Fig.~\ref{Fig11}), where we previously observed that the original large-scale dynamics of $T_{\textrm{max}}$ (OTS) has a much higher value of statistical complexity $C$ and a lower value of permutation entropy $H$. One of the factors that can increase complexity could be temporal nonlinear dependencies, such as long-range correlations over large-scale dynamics or multiscale fractal scaling properties. Extension of our hybrid model~\eqref{eq:hm} incorporating such long-range dynamics is a future outlook of our present work.

Additionally, in Fig.~\ref{Fig14}, we plot the annual maxima and minima of the $\hat{T}_{\textrm{max}}$ time series from the three models of WN(L), CN(L), and LN(L) (previously seen in Fig.~\ref{Fig12}). Each model is extrapolated to highlight the predicted increasing trends in $\hat{T}_{\textrm{max}}$. The vertical dashed line indicates the current year of 2025. There is a sharp rise in $\hat{T}_{\textrm{max}}$ starting in the period from 2003-2007. In the predicted regime, both the maxima and minima of $\hat{T}_{\textrm{max}}$ reach values significantly higher than those recorded in the previous years. Our hybrid modeling framework thus implies a concerning trend in maximum temperature dynamics of Imphal, thereby highlighting the need for the implementation of effective climate control measures to mitigate this warming trend. Our observation aligns with the deforestation records of Manipur state, where 255 kha of tree cover has been recorded as lost from the period from 2001-2024~\cite{GFW_Manipur}.

To quantify the predictive skill of our hybrid model, we perform hindcast experiments by re-initializing the model at past years, 1990 and 2010, and comparing the simulated annual temperature maxima and minima with the observed data, as shown in Fig.~\ref{Fig15}. We simulate the model for two periods - two decades from 1990 and one decade from 2010 - and evaluate the performance of the model using $RMSE$ and Pearson correlation (results shown in Table~\ref{Table2}). The hindcasts show good agreement with the observed extremes, with most observations lying within the 95\% confidence intervals. We find that the $RMSE$ values are slightly larger for annual minima despite higher correlations, which might be caused by unresolved nocturnal and radiative processes~\cite{taylor2001} not explicitly represented in the model.

\renewcommand{\arraystretch}{1.3}
\begin{table*}
\centering
\caption{\label{Table2}Hindcast skill metrics for annual temperature maxima and minima using different noise formulations}
\begin{tabular}{|c|c|cc|cc|}
\hline
\textbf{Noise} & \textbf{Period} & \multicolumn{2}{c|}{\textbf{Maxima}} & \multicolumn{2}{c|}{\textbf{Minima}} \\
\cline{3-4} \cline{5-6}
 &  & $RMSE$ ($^\circ$C) & $r$ & $RMSE$ ($^\circ$C) & $r$ \\
\hline
White   & 1990-2010 & 0.790 & 0.447 & 1.065 & 0.719 \\
        & 2010-2020 & 0.459 & 0.686 & 2.089 & 0.938 \\
\hline
Colored & 1990-2010 & 0.803 & 0.547 & 1.082 & 0.717 \\
        & 2010-2020 & 0.645 & 0.628 & 2.404 & 0.935 \\
\hline
L\'evy  & 1990-2010 & 0.777 & 0.527 & 1.105 & 0.680 \\
        & 2010-2020 & 0.433 & 0.595 & 1.834 & 0.924 \\
\hline
\end{tabular}
\vspace{0.5em}
\begin{center}
$RMSE$: Root Mean Square Error;
$r$: Pearson Correlation Coefficient.
\end{center}
\end{table*}
\renewcommand{\arraystretch}{1.0}

\subsection{Langevin \& Fokker-Planck Equations for $\hat{T}_{\textrm{max}}$}
We now derive the Langevin and Fokker-Planck equations for the hybrid model of $\hat{T}_{\textrm{max}}$ dynamics (Eq.~\eqref{fin_model}) incorporating the three noise types (white, colored, and Lévy) considered in the preceding analyses.

To get the Langevin equation of $\hat{T}_{\textrm{max}}$, differentiating Eq.~\eqref{fin_model} with time, in It\^o convention, yields:
\begin{eqnarray}
    \label{Langevin_gen}
    \frac{d\hat{T}_{\mathrm{max}}(t)}{dt}=\psi(t)+\Gamma\frac{d\zeta(t)}{dt},
\end{eqnarray}
where $\psi(t)=\displaystyle\frac{d\Lambda(t)}{dt}+\frac{dF(t)}{dt}$ is the drift term and $\displaystyle\Gamma\frac{d\zeta(t)}{dt}$ represents the diffusion term.
Here, $\Lambda(t)$
is the deterministic component of Eq.~\eqref{deterministic} and $F(t)$ is the nonlinear feedback term of Eq.~\eqref{cubic_fb} or \eqref{Lorenz_fb}. 

When $\zeta(t)$ is modeled as a standard Wiener process, its derivative $\displaystyle\frac{d\zeta(t)}{dt}$ corresponds to Gaussian white noise $\eta(t)$ with zero mean and delta-correlated fluctuations as $\langle\eta(t)\eta(t^\prime)\rangle=\delta(t-t^\prime)$. The Langevin equation then becomes:
\begin{eqnarray}
    \label{Langevin_white}
    \frac{d\hat{T}_{\mathrm{max}}(t)}{dt}=\psi(t)+\Gamma\eta(t).
\end{eqnarray}
The associated Fokker-Planck equation~\cite{Risken96,Gardiner85} for the time evolution of the probability density of $\hat{T}_{\mathrm{max}}$, denoted by $P(\hat{T}_{\mathrm{max}},t)$ is
\small{
\begin{eqnarray}
    \label{FP_white}
    \frac{\partial P(\hat{T}_{\mathrm{max}},t)}{\partial t}=-\frac{\partial}{\partial\hat{T}_{\mathrm{max}}}[\psi(t)P(\hat{T}_{\mathrm{max}},t)]\nonumber\\
    +\frac{\Gamma^2}{2}\frac{\partial^2P(\hat{T}_{\mathrm{max}},t)}{\partial\hat{T}_{\mathrm{max}}^2}.
\end{eqnarray}
}
\normalsize
For colored noise, $\zeta(t)$ is a correlated stochastic process with noise correlation $\langle\eta(t)\eta(t^\prime)\rangle=\kappa(t-t^\prime)$, where $\kappa(t-t^\prime)$ is a smooth, decaying correlation function of time difference. It is typically modeled as an Ornstein-Uhlenbeck (OU) process, which is an exponentially correlated noise with a correlation time $\tau$, given by~\cite{Giuggioli19,Van89,Haunggi94,Lien2025,Alexandrov2025}:
\begin{eqnarray}
    \label{color}
    \frac{d\zeta(t)}{dt}=-\frac{1}{\tau}\zeta(t)+\sqrt{\frac{2D}{\tau}}\eta(t), \label{ou}
\end{eqnarray}
where 
$D$ is the noise strength. Substituting this Eq.~\eqref{ou} into the Langevin equation~\eqref{Langevin_gen}, we get a coupled set of stochastic differential equations. The corresponding Fokker-Planck equation~\cite{Giuggioli19,Van89,Haunggi94,Lien2025,Alexandrov2025} that describes the time evolution of the joint probability density function of $\hat{T}_{\mathrm{max}}$ and $\zeta(t)$ is:
\begin{align}
    \label{FP_color}
    \frac{\partial P(\hat{T}_{\mathrm{max}},\zeta,t)}{\partial t}&=-\frac{\partial}{\partial\hat{T}_{\mathrm{max}}}[\lbrace\psi(t)+\Gamma\zeta(t)\rbrace P(\hat{T}_{\mathrm{max}},\zeta,t)]\nonumber\\
    &+\frac{\partial}{\partial\zeta}\left[\frac{\zeta}{\tau}P(\hat{T}_{\mathrm{max}},\zeta,t)\right]\nonumber\\
    &+D\frac{\partial^2 P(\hat{T}_{\mathrm{max}},\zeta,t)}{\partial\zeta^2}.
\end{align}

In the case of Lévy noise, the Langevin equation becomes~\cite{Denisov09,Zan20,Chechkin06,Metzler00,Zheng2020}:
\begin{eqnarray}
    \label{Langevin_levy}
    \frac{d\hat{T}_{\mathrm{max}}(t)}{dt}=\psi(t)+\Gamma L(t),
\end{eqnarray}
where $L(t)=\displaystyle\frac{d\zeta(t)}{dt}$ is a Lévy process with stability index $\alpha$ (where $0<\alpha<2$). The corresponding Fokker-Planck equation generalizes to a fractional Fokker-Planck equation~\cite{Denisov09,Zan20,Chechkin06,Metzler00,Zheng2020} as:
\small{
\begin{align}
    \label{FP_levy}
    \frac{\partial P(\hat{T}_{\mathrm{max}},\zeta,t)}{\partial t}=-\frac{\partial}{\partial\hat{T}_{\mathrm{max}}}[\psi(t) P(\hat{T}_{\mathrm{max}},\zeta,t)]\nonumber\\
    +\Gamma^\alpha\frac{\partial^\alpha P(\hat{T}_{\mathrm{max}},\zeta,t)}{\partial|\hat{T}_{\mathrm{max}}|^\alpha},
\end{align}
}
\normalsize where $\displaystyle\frac{\partial^\alpha}{\partial|\hat{T}_{\mathrm{max}}|^\alpha}$ is the Riesz fractional derivative~\cite{Metzler00,Samko93,Chechkin02}, which captures the non-local, jump-driven nature of the Lévy process.

These formulations offer a theoretical basis that links the hybrid model of maximum temperature $\hat{T}_{\textrm{max}}$ dynamics (Eq.~\eqref{fin_model}) to underlying physical and statistical principles. While the Langevin equations of $\hat{T}_{\textrm{max}}$ capture its evolution under deterministic dynamics, feedback, and stochastic noise, the corresponding Fokker-Planck equations describe the evolution of the probability distribution of $\hat{T}_{\textrm{max}}$ or its joint probability over time. These derivations of Langevin and Fokker-Planck equations of $\hat{T}_{\textrm{max}}$ indicate how different noise types influence the temporal dynamics or probability distributions of $\hat{T}_{\textrm{max}}$, providing a theoretical understanding of maximum temperature dynamics. We thus establish a general probabilistic framework that can help in a deeper theoretical understanding and analytical interpretation of maximum temperature dynamics.

\section{Conclusion}
\label{sec:conc}
In this work, we introduce a comprehensive hybrid dynamical-stochastic framework to model the variability of maximum temperature observed in the capital city Imphal of Manipur, located in Northeast India. The framework combines deterministic components (derived from spectral decomposition and Fourier analysis) with stochastic elements, including white, colored, and Lévy noise. It also incorporates a nonlinear feedback mechanism through a cubic term and a feedback term derived from the Lorenz system, capturing chaotic influences. The model therefore reflects both quasi-periodic deterministic trends and random fluctuations observed in the empirical maximum temperature time series data. For our analysis, we have used publicly available data on the maximum temperature records of Imphal spanning a duration of 73 years. 

Our findings show that the hybrid modeling approach, which combines deterministic spectral analysis with statistically informed noise diagnostics and chaotic feedback, effectively models the observed maximum temperature dynamics time series at the regional scale. Statistical tests indicate that our hybrid model has good accuracy with respect to the empirical time series data when a monthly timescale is used. Further validation using the nonlinear analysis technique of the complexity-entropy ($CH$) causality plane shows that the hybrid model, with different feedback and appropriate noise terms, reproduces the entropy and complexity characteristics of the original time series data better than purely deterministic models at the monthly timescale. Among the feedback mechanisms tested, the $(H,C)$ values of the Lorenz feedback match better with the observed data, implying similar degrees of entropy and complexity. However, our results show that the hybrid model becomes insufficient when considering large-scale temporal dynamics over years, and suggest the inclusion of higher-order terms in the model for better modeling of the observed dynamics.

We further formulate a theoretical foundation for our hybrid model of maximum temperature dynamics by deriving the associated Langevin and Fokker-Planck equations using different noise types. These derivations show how the interplay of deterministic drift, feedback, and stochastic diffusion leads to the temporal evolution of maximum temperature and how it shapes the probability distribution of maximum temperature. This formulation establishes a general probabilistic framework that can help in a deeper theoretical understanding and analytical interpretation of the dynamics of regional climate patterns.

Further extension of the hybrid model can include: (i) adaptive feedback parameters that enables the system to adjust to evolving warming trends; (ii) multi-scale decomposition methods such as empirical mode decomposition (EMD); and (iii) coupling with physical climate predictors such as precipitation, soil moisture, and forest cover to constrain long-term dynamics through observed interactions. Incorporating these extensions will make the hybrid model more robust and enhance its predictive performance.

The hybrid modeling framework introduced in this work serves as a preliminary study aimed at understanding the ongoing impacts of climate change in the state of Manipur, with a particular focus on the dynamics of maximum temperature. Since the different components of the hybrid model capture various aspects of the temperature dynamics (deterministic trends reflect seasonal cycles or anthropogenic forcing, feedback mechanisms represent nonlinear responses, and the noise term captures variability), one can analyze and compare the relative contributions of these components. This will allow us to identify and quantify the contributing climatic factors and primary causes of the observed temperature rise in the region. Such an investigation will provide an in-depth assessment of the underlying physical, environmental, and possibly the socio-economic factors contributing to regional climate change in Manipur. Our hybrid modeling framework is generalizable to other climatic variables and geographic regions.

Beyond its immediate application to climate modeling, the hybrid modeling approach proposed in this study holds potential for extension to time series analysis in other fields such as neuroscience, ecology, and geophysics, where both deterministic cycles and stochastic influences coexist. Its ability to decompose, reconstruct, and characterize signals across scales makes it well-suited for detecting anomalies in the dynamics of complex systems.

\begin{acknowledgments}
M.K.S. is partly supported by the National Fellowship for Scheduled Castes Students (NFSC), provided by the National Scheduled Castes Finance and Development Corporation (NSFDC) and the Department of Social Justice \& Empowerment, Ministry of Social Justice \& Empowerment, Government of India, with reference number 201610028460. M.K.S. would like to thank the North American Manipur Association (NAMA) for the Dr. A. Surjalal Sharma Memorial Grant award. A.L.C. acknowledges the APCTP (JRG program) through the Science and Technology Promotion Fund and Lottery Fund of the Korean Government and the Korean Local governments-Gyeongsangbuk-do Province and Pohang City. In addition, the authors would like to thank Samananda Keisham, Department of Forestry \& Environmental Science, Manipur University, for the insightful discussions, which contributed to the development of this work.
\end{acknowledgments}

\section*{Author Declarations}
\subsection*{Conflict of Interest}
The authors have no conflicts to disclose.

\subsection*{Author Contributions}
\textbf{M.K.S.}: Conceptualization (lead); Methodology (lead); Data curation (lead); Formal analysis (lead); Software (lead); Validation (equal); Visualization (equal); Writing – original draft (equal). \textbf{A.L.C.}: Formal analysis (supporting); Validation (equal); Visualization (equal); Writing – original draft (equal). \textbf{R.K.B.S.}: Conceptualization (supporting); Supervision (equal); Writing – review and editing (equal). \textbf{M.S.S.}: Conceptualization (supporting); Supervision (equal); Writing – review and editing (equal).

\section*{Data Availability}
Data available on request from the authors.

\appendix
\section{Kalman Filter}
\label{sec:kf}
The Kalman filter is a recursive Bayesian estimator that updates predictions of a system's state based on both prior estimates and incoming observations~\cite{Durbin12,Shumway00,Adejumo2021}. In our implementation, the filter is initialized with the first observation as the starting estimate of the hidden state. We then employ the Expectation-Maximization (EM) algorithm over 10 iterations to estimate optimal model parameters. Next, we apply the Kalman smoother to incorporate both past and future observations, which can produce a more stable and accurate estimate of the latent signal. Finally, the two-dimensional smoothed output is then flattened into a one-dimensional array for subsequent time series analyses.

\section{Noise Diagnostics}
\label{sec:nd}
\renewcommand{\thefigure}{B\arabic{figure}}
\setcounter{figure}{0}

The diagnostics of the noise component consist of four parts as follows:
\begin{itemize}
    \item \textit{Kernel Density Estimation (KDE)}: It is a non-parametric method used to estimate the probability density function (PDF) of a random variable~\cite{Scott15,Silverman18}. It works by placing a smooth ``kernel" function (commonly a Gaussian bell-shaped curve) at each data point and then summing these kernels to form the overall density estimate. We first evaluate the mean and standard deviation of the noise component, and then the KDE function approximates the PDF of the noise component to analyze whether the data is normally distributed, skewed, or heavy-tailed. The bandwidth is selected using the Scott's rule~\cite{Scott15} and cross-validated using Silverman's rule~\cite{Silverman18} to balance bias–variance trade-off and ensure smooth yet accurate probability density estimation.
    \item \textit{Stability and Skewness parameters estimation}: The stability parameter $\alpha$ of a Lévy alpha-stable distribution controls the tail thickness. A null skewness parameter implies a symmetric distribution. We fit the Lévy alpha-stable distribution to the noise component, where the best-fitting parameters are determined using maximum likelihood estimation~\cite{Samoro94,Nolan20,Nolan97}.
    \item \textit{Power Spectral Density (PSD) and Spectral Decay parameter estimation}: By computing the PSD using Welch's method~\cite{Welch03}, we quantify the spectral structure of the noise component. We perform a linear regression on the log–log plot of frequency vs.\ PSD, where the slope of this regression line (multiplied by –1) yields the spectral decay parameter $\beta$. The noise spectrum follows the relation, $\displaystyle PSD(\nu)=\frac{1}{\nu^\beta}$, where the value of $\beta$ classifies the noise type and provides a quantitative metric for understanding the memory and correlation structure in the noise component~\cite{Voss78,Mandelbrot82,Milotti02}. 
    \begin{figure}
    \includegraphics[scale=0.3]{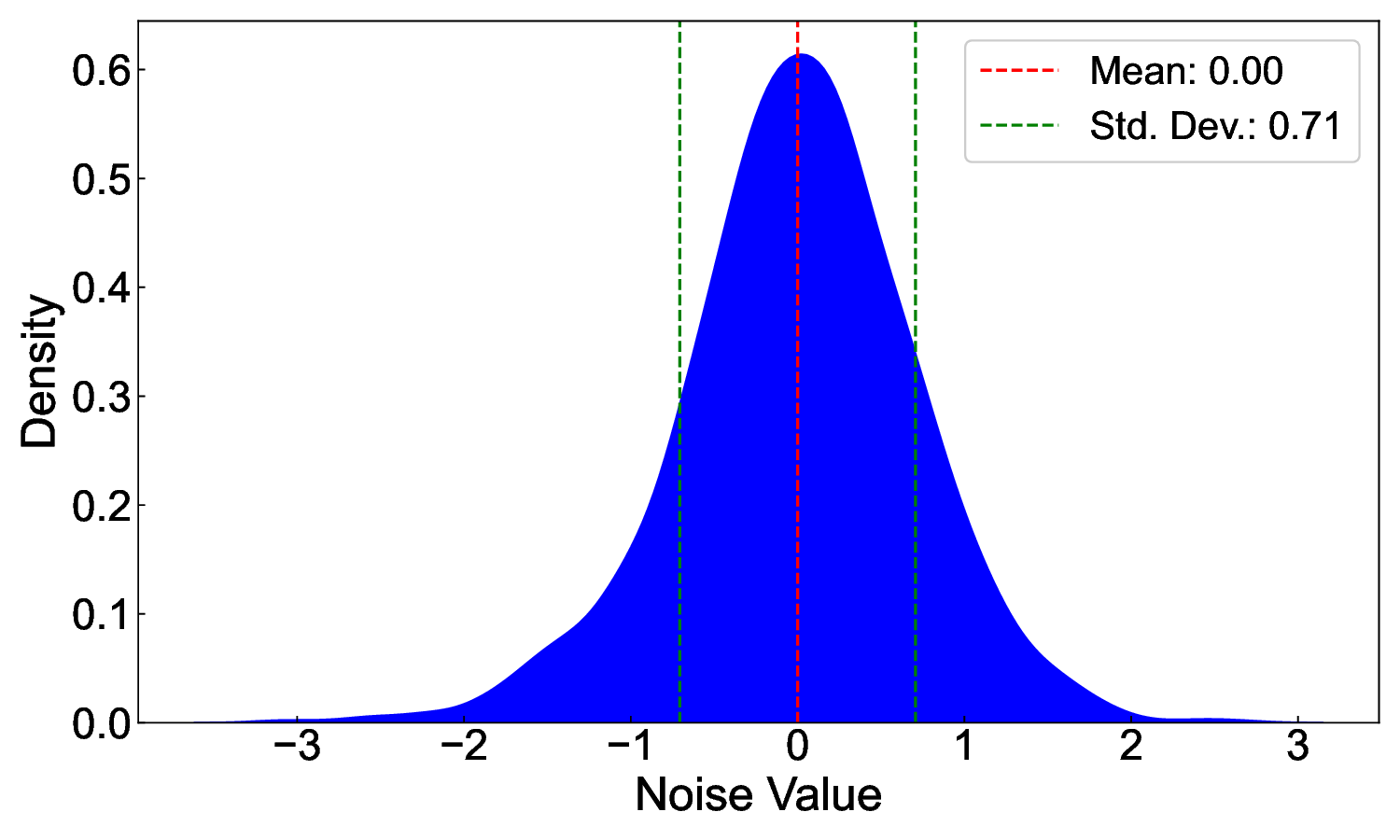}
    \caption{Plot of the probability density function (PDF) of the noise component obtained from singular spectral analysis (SSA) for the monthly data of January of the period 1951-2024. Mean and standard deviation are indicated by vertical lines (see colour in the legend).}
    \label{FigB1}
    \end{figure}
    \begin{figure}
    \includegraphics[scale=0.4]{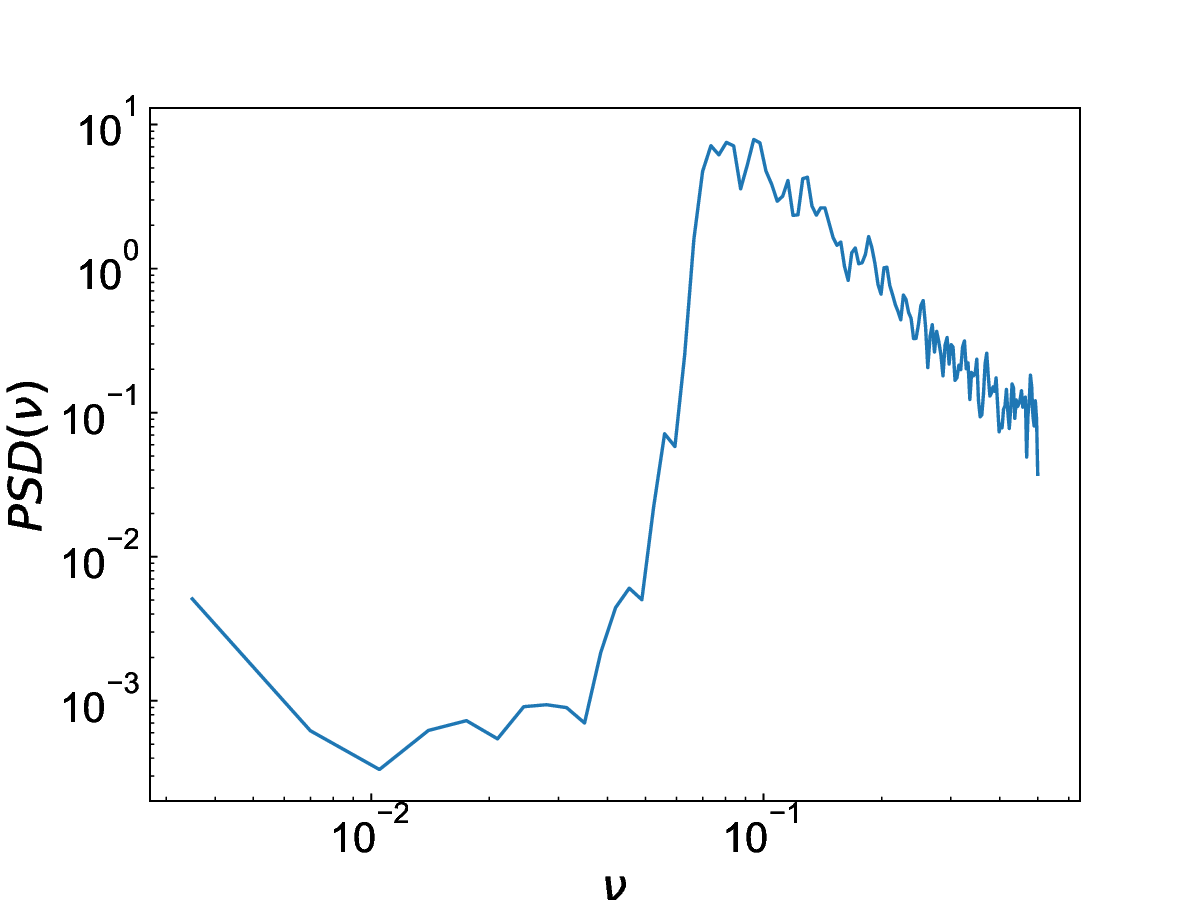}
    \caption{Power spectrum density (PSD) plot of the noise component obtained from singular spectral analysis (SSA) for the monthly data of January of the period 1951-2024. Both axes are on a log scale. We can see that the log-power increases at small frequencies but decreases at larger frequencies. Overall, a linear regression fit (not shown here) shows a positive slope.}
    \label{FigB2}
    \end{figure}
    \begin{figure}
    \includegraphics[scale=0.35]{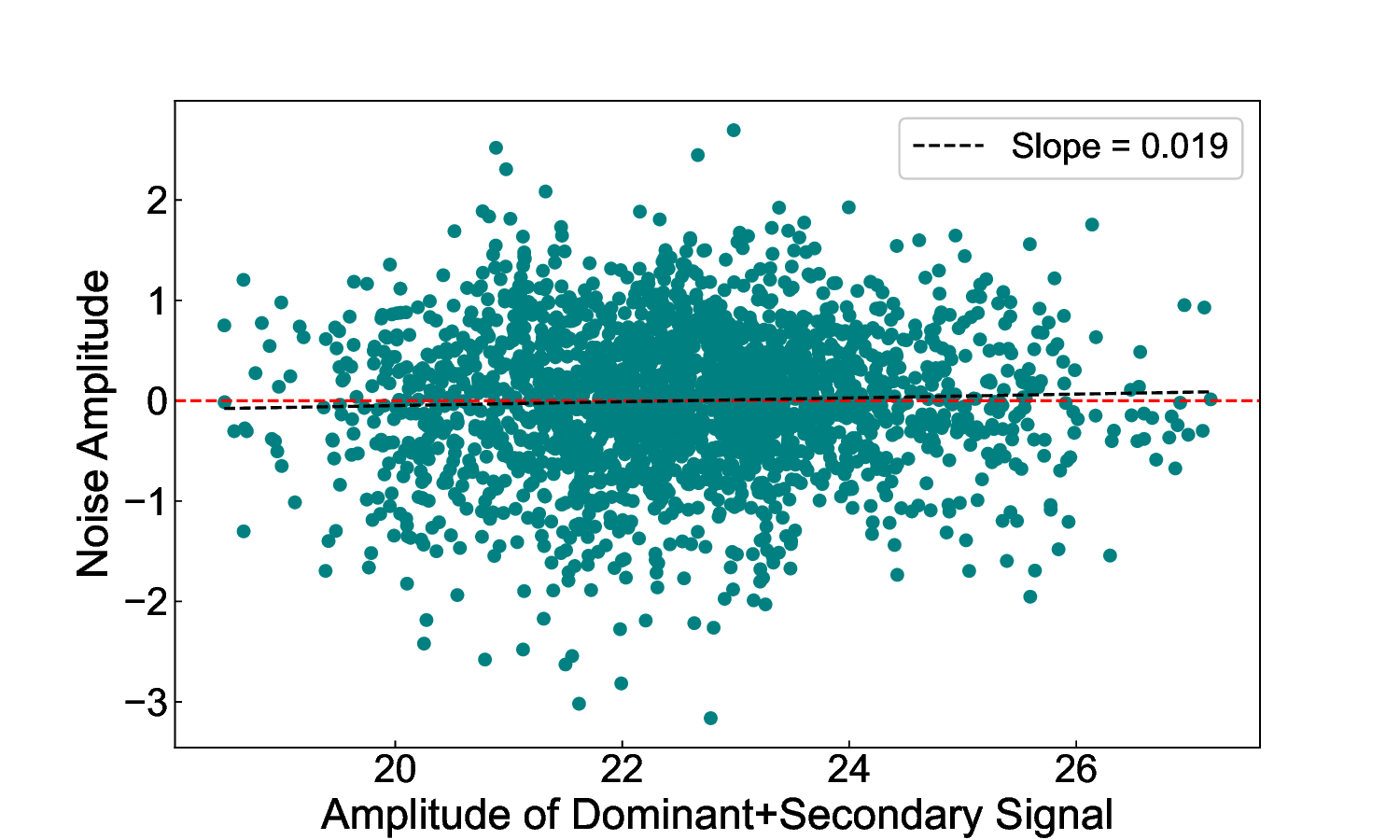}
    \caption{Scatter plot of the amplitude of the noise component against the signal amplitude of the dominant and secondary components for the monthly data of January of the period 1951-2024. The dashed black line indicates the linear regression fit with its slope = 0.019 indicated in the legend. The dashed red line is simply a visual guideline at zero noise amplitude.}
    \label{FigB3}
    \end{figure}
    \item \textit{Additive/Multiplicative test}: We perform a test to identify whether the noise component is additive or multiplicative by plotting a scatter plot of the noise component against the signal (the sum of dominant and secondary components extracted via SSA). Additive noise is independent of the signal amplitude, whereas multiplicative noise scales with the signal's amplitude~\cite{Allen96,Beran17,Kantz03,Gardiner85}. If the noise amplitude shows no correlation with the signal and displays a flat or near-horizontal distribution, it suggests additive behavior. Conversely, an increasing or decreasing trend implies a signal-dependent or multiplicative noise structure. 
\end{itemize}

\section{Statistical and complexity tests}
\label{sec:at}
\begin{itemize}
    \item \textit{Statistical metrics}: To determine the modeling accuracy of our hybrid model with respect to the original time series, we compute three metrics, namely, the coefficient of determination ($R^2$)~\cite{Wright21,Ozer85}, Root Mean Square Error ($RMSE$)~\cite{Armstrong92} and Kling-Gupta Efficiency ($KGE$)~\cite{Gupta11} defined as follows:
    \begin{eqnarray}
    \label{stat_tests}
    \left.
    \begin{array}{l}
   R^2=1-\frac{\displaystyle\sum_{i=1}^n (T^i_{\mathrm{max}}-\hat{T}^i_{\mathrm{max}})^2}{\displaystyle\sum_{i=1}^n (T^i_{\mathrm{max}}-\bar{T}^i_{\mathrm{max}})^2},\\
    RMSE=\sqrt{\displaystyle\frac{1}{n}\displaystyle\sum_{i=1}^n (T^i_{\mathrm{max}}-\hat{T}^i_{\mathrm{max}})^2},\\
    KGE=1-\sqrt{(r-1)^2+(\theta-1)^2+(\gamma-1)^2}.
    \end{array}
    \right\}
    \end{eqnarray}
    Here, $T_{\mathrm{max}}$ and $\hat{T}_{\mathrm{max}}$ respectively represent the maximum temperature values from the original time series data and the model, and $\bar{T}_{\mathrm{max}}$ is the corresponding mean from the original data. In the definition of KGE, $r$ denotes the Pearson correlation coefficient between the original and model data, $\displaystyle\theta\left(=\frac{\bar{\hat{T}}_{\mathrm{max}}}{\bar{T}_{\mathrm{max}}}\right)$ is the bias ratio with the mean of the model data $\bar{\hat{T}}_{\mathrm{max}}$, and $\gamma$ is the ratio of coefficient of variation of the model to the original data. $n$ denotes the total number of observations.
    \item \textit{Validation with nonlinear analysis using complexity–entropy ($CH$) plane}: The statistical complexity $C$ is defined as:
    \begin{eqnarray}
    \label{complexityy}
    C=D_E\times S,
    \end{eqnarray}
   where $D_E$ denotes \textit{disequilibrium} (how far the system's probability distribution is from a uniform distribution) and $S\left(=-\displaystyle \sum_{i}p_i\log_2(p_i)\right)$ is the Shannon entropy.
   
  Bandt and Pompe~\cite{Bandt02} proposed an entropy known as permutation entropy, based on Shannon entropy, which quantifies the degree of disorder in a given time series based on ordinal patterns. The $ p_i$'s in the definition of Shannon entropy then become the relative frequency of observing $i$ type of ordinal patterns~\cite{Bandt02} in the given time series using the embedding dimension $d$. Normalized permutation entropy is defined as:  
    \begin{eqnarray}
    \label{shannon}
    H=-\sum_{i=1}^{d!}\frac{S}{\log_2(d!)},
    \end{eqnarray}
     
    A common disequilibrium quantity used is the Jensen–Shannon divergence (denoted by $Q_J$). Connecting with $H$, the statistical complexity $C$ is redefined as:
    \begin{eqnarray}
    \label{complexity}
    C=Q_J\times H.
    \end{eqnarray}
   $Q_J$ measures the distance of the observed distribution of ordinal patterns ($P$) from a uniform distribution ($P_e$) as:
    \begin{eqnarray}
    \label{JSD}
    Q_J(P, P_e)=Q_0\left[S\left(\frac{P+P_e}{2}\right)-\frac{1}{2}S(P)\right.\nonumber\\
    \left.-\frac{1}{2}S(P_e)\right],
    \end{eqnarray}
    where $Q_0$ is a normalization constant.

    The complexity–entropy ($CH$) causality plane~\cite{Rosso07,Zunino12,Huang2021} is a two-dimensional representation of permutation entropy ($H$) and statistical complexity ($C$). Purely random processes tend to have high entropy and low complexity, while chaotic systems occupy regions with moderate entropy and high complexity~\cite{Rosso07}. The \textit{CH}-plane offers a model-free, data-driven approach that is robust even for short and noisy time series~\cite{Martin06}.
\end{itemize}

\bibliography{References}

\end{document}